\documentclass[twocolumn,showpacs,showkeys,prd,preprintnumbers,amsmath,amssymb]{revtex4}

\usepackage[]{epsfig,dcolumn}
\usepackage{graphicx}
\usepackage{dcolumn}
\usepackage{bm}

\begin{document}


\title{A partial wave analysis of the $\pi^- \pi^- \pi^+$ and $\pi^- \pi^0 \pi^0$ systems\\
and the search for a $J^{PC}=1^{-+}$ meson}
\author{A.~R.~Dzierba}
\author{R.~Mitchell}
\author{E.~Scott}
\author{P.~Smith}
\author{M.~Swat}
\author{S.~Teige}
\affiliation{Department of Physics, Indiana University, Bloomington, IN 47405}
\author{A.~P.~Szczepaniak}
\affiliation{Department of Physics and Nuclear Theory Center, 
 Indiana University, Bloomington, IN 47405}
 \author{S.~P.~Denisov}
 \author{V.~ Dorofeev}
 \author{I.~ Kachaev}
  \author{V.~ Lipaev}
 \author{A.~V.~Popov }
 \author{D.~I.~Ryabchikov}
 \affiliation{Institute for High Energy Physics, Protvino, Russian Federation 142284}
  \author{V.~A.~Bodyagin }
 \author{A.~Demianov}
 \affiliation{Nuclear Physics Institute, Moscow State University, Moscow, Russian Federation 119992}
\date{\today}

\date{\today}

\begin{abstract}
A partial wave analysis (PWA) of the  $\pi^- \pi^- \pi^+$ and $\pi^- \pi^0 \pi^0$ systems 
produced in the reaction $\pi^- p \to (3\pi)^-p$ at 18~GeV/$c$ was carried
out using an \emph{isobar} model assumption.  This analysis is based on 3.0M
$\pi^- \pi^0 \pi^0$ events and 2.6M  $\pi^- \pi^- \pi^+$  events and
shows production
of the $a_1(1260)$, $a_2(1320)$, $\pi_2(1670)$
and $a_4(2040)$  resonances.   
Results of detailed studies of the stability
of partial wave fits are presented. 
An earlier analysis of 
250K $\pi^- \pi^- \pi^+$ events from the same experiment
 showed possible evidence for a $J^{PC}=1^{-+}$ exotic meson
with a mass of $\sim$1.6~GeV/$c^2$ decaying into $\rho \pi$.
In this analysis of a higher statistics sample of the $(3\pi)^-$ system in two
charged modes
we find no evidence of an exotic meson.
\end{abstract}

\pacs{11.80.Cr, 13.60.Le, 13.60Rj}
\keywords{meson resonances}
\maketitle

\section{\label{sec:intro}Introduction}

In this paper we present a partial wave analysis (PWA) of a
high-statistics sample of events corresponding to the production of the
 $(3\pi)^{-}$ system produced in $\pi^-p$ collisions in two modes:  $\pi^- \pi^0\pi^0 $ 
and $ \pi^- \pi^- \pi^+ $.  This sample size exceeds, by at least an order of
magnitude, the largest published sample size of $3\pi$ events to date. 
 Previous
$3\pi$ analyses led to the discovery and/or determination of properties 
 of the $a_1(1260)$, $a_2(1320)$, 
$\pi_2(1670)$,  $\pi(1800)$ and the $a_4(2040)$ resonances
\cite{Ballam68,Ascoli70,Antipov73,Baltay77,Daum80a,Daum80b,Daum80c,Daum81}.  

In 1998, the E852 collaboration reported evidence for the $\pi_1(1600)$, a 
 $J^{PC}=1^{-+}$ exotic hybrid meson with a mass of 1.6~GeV/$c^2$ decaying
 into $\rho \pi$  \cite{Adams98,Chung02}.  
 That analysis was  based on 250,000 events 
 of the reaction $\pi^- p \to \pi^- \pi^- \pi^+ p$ collected in 1994. 
We report on the analysis of additional data collected in
 1995 including 3.0M events of the 
 reaction   $\pi^- p \to \pi^- \pi^0\pi^0 p$ and 2.6M events of the reaction
  $\pi^- p \to \pi^- \pi^- \pi^+ p$.

The identification of exotic mesons requires a PWA to extract  signals.
The $3\pi$ system provides a particularly attractive venue for such
searches since earlier work suggests a rich spectrum of meson resonances.
If an exotic meson is produced with relatively small amplitude, its
interference with nearby well-established resonances is a sensitive
search tool.

In almost all of the published amplitude analyses of the $3\pi$ system, 
the \emph{isobar}
model was employed -- a $3\pi$ system 
with a particular $J^{PC}$ is produced and  decays into a  di-pion resonance
with well-defined quantum numbers and a bachelor $\pi$ followed by the decay of
the di-pion resonance.  This assumption is successful in describing many
features of the $3\pi$ system and is motivated by the 
observation that the di-pion effective mass spectrum shows
prominent resonance production, for example, the 
$\rho(770)$ and the $f_2(1275)$. The di-pion resonances
considered in this analysis include the 
 $f_0(980)$, $\rho(770)$,
 $f_2(1275)$ and the $\rho_3(1690)$.  We also include parametrizations
 of $S$-wave $\pi\pi$ scattering.

 In this analysis,  we observe the 
 $a_1(1260)$, $a_2(1320)$, 
$\pi_2(1670)$ and the $a_4(2040)$ in appropriate partial
wave intensities and in their relative phase differences.  We also searched for,
but find no evidence for, the  $J^{PC}=1^{-+}$ exotic   $\pi_1(1600)$ in either
the   $\pi^- \pi^0\pi^0 $ or  $ \pi^- \pi^- \pi^+ $ mode.  

In order to extract reliable information from a partial wave analysis it
is important to establish a procedure for determining a sufficient set of partial
waves.  Failure to include important partial waves 
in the series expansion may
lead to inconsistent  results
and erroneous conclusions.  In this paper we describe our procedure for determining
a sufficient wave set.  That
procedure includes a comparison of moments as calculated from
PWA solutions with those computed directly from data.  
We emphasize that this analysis 
is similar to that of  references~ \cite{Adams98,Chung02} in that the same
isobar model assumptions are made but the final set of partial waves used
is different.  Both analyses make the same assumptions about coherence
between different partial waves.  It is possible that relaxing these
coherence assumptions could lead to different results in both analyses.

\subsection{General features of the $\bm{(3\pi)^-}$ system}

The  $3\pi$ system with non-zero charge has isospin $I > 0$, and, since
 no flavor exotic mesons have been found, we assume $I=1$. Since
a state with an odd number of pions has negative $G$~parity, the
relationship $G=C(-1)^I$
implies positive $C$~parity for the $(3\pi)^-$ system.

The simultaneous observation of the  two  $(3\pi)^-$ modes,  $\pi^- \pi^0\pi^0 $ 
and $ \pi^- \pi^- \pi^+ $,  in the same experiment provides important cross
checks.  Consider the production of resonance $X$ and its subsequent decay into
$(3\pi)^-$ via an intermediate di-pion resonance.
  If the intermediate di-pion resonance 
is an isoscalar, then the yield into $ \pi^- \pi^- \pi^+ $ should be
twice that into $\pi^- \pi^0\pi^0 $.  Similarly, if the di-pion is an isovector, then
there should be equal yields into $ \pi^- \pi^- \pi^+ $ and $\pi^- \pi^0\pi^0 $.  Since
the $\pi^- \pi^- \pi^+$ and $\pi^- \pi^0 \pi^0$ modes rely on different elements of the
detector,  any misunderstandings in the acceptance would affect the two
modes differently and ultimately lead to inconsistent interpretations of the PWA results.
This underscores
the importance of having these two different modes to validate the analysis.

\subsection{Exotic mesons within QCD}

Quantum chromodynamics (QCD) predicts a spectrum of mesons beyond the
$q \bar q$ bound states of the conventional quark model.
 The spin ($J$), parity ($P$) and charge conjugation ($C$) 
 quantum numbers of
a $q \bar q$ system are: $\vec{J}=\vec{L}+\vec{S}$
where $\vec{L}$ is the angular momentum between the quarks and $\vec{S}$ is
the total quark spin,  $P=(-1)^{L+1}$ and $C=(-1)^{L+S}$.  The $J^{PC}$
combinations: $0^{+-},
1^{-+}, 2^{+-}, \dots$ are not allowed and are called {\it exotic} .  
Lattice QCD \cite{Bali2004}
and QCD-inspired models predict that the excitation of the
gluonic field within a meson leads to {\it hybrid} mesons, where the 
gluonic degrees of freedom allow hybrid mesons to have 
exotic quantum numbers.  The observation of hybrid mesons and measurement
of their properties provides experimental input necessary for an understanding
of quark and gluon confinement in QCD.

\subsection{Organization of this paper}

This paper is organized as follows.  Section~\ref{sec:experiment} is a discussion
of experimental details, including a description of the apparatus (\ref{sec:apparatus}),
event reconstruction and selections (\ref{sec:data}) and distributions in 
effective mass before and after acceptance corrections (\ref{sec:masses}).
An overview of the PWA methodology, including selection of the wave set, is given in
Section~\ref{sec:pwamethod}.  Section~\ref{sec:pwaresults} is a presentation of the PWA results
for individual partial waves using the wave set used in this analysis and the wave
set used in the analysis of references \cite{Adams98,Chung02} for both the
$\pi^-\pi^0\pi^0$ and $\pi^-\pi^-\pi^+$ modes.  Section~\ref{sec:exotica} focuses on the 
exotic $J^{PC}=1^{-+}$ waves and how the failure to include important 
$J^{PC}=2^{-+}$ waves could lead to misleading conclusions regarding the
existence of an exotic resonance.  Section~\ref{sec:systematics} discusses studies of systematic
uncertainties and the conclusions are presented in Section~\ref{sec:conclude}.

\section{\label{sec:experiment}Experimental details}

\subsection{\label{sec:apparatus}The experimental apparatus}

 \begin{figure}  
\centerline{\epsfig{file= 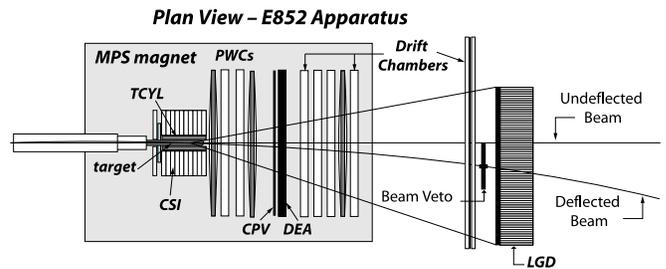,width=\columnwidth}}
\caption{
E852 apparatus and the MPS at BNL.
}\label{MPS}
\end{figure}

The E852 experiment used the upgraded multiparticle spectrometer (MPS) at the 
Alternating Gradient Synchrotron (AGS) at Brookhaven National Lab (BNL). 
The apparatus is described in detail elsewhere
\cite{Teige99} and is shown schematically in Figure~\ref{MPS}.  
 The MPS
was upgraded for E852 with the inclusion of a cylindrical drift chamber (TCYL) \cite{Bar-yam97}
and a cylindrical cesium iodide (CSI) 
detector \cite{AdamsT96}, both concentric with a  LH$_2$
target.  Downstream of the target were 
 proportional wire chambers (PWCs) for triggering and tracking,
 a lead/scintillator array (DEA) and a 3000-element
lead glass electromagnetic calorimeter (LGD). 
The CSI calorimeter consisted of 180 segments -- 10 longitudinal segments
and 18 azimuthal segments.  The DEA had the geometry of a picture
frame with a rectangular central hole.  A scintillation counter (CPV) was
placed just upstream of and matched the geometry of the DEA.
The lead glass blocks of the LGD
\cite{Brabson93,Crittenden97,Gunter97,Lindenbusch97}
 were each 45~cm long with a transverse
area of $4\times4$~cm$^2$ and were arranged in a  $284\times172$~cm$^2$ stack
with a 2~block $\times$ 2~block beam hole to allow for passage of
non-interacting beam particles.
 The DEA and CSI detected low-energy
photons that miss the LGD.
The LGD was used to measure energies and positions of photons
from $\pi^0$ and $\eta$ mesons.  The MPS included an analyzing magnet
with a 1~T field  and 
drift chambers for tracking. The LH$_2$ target was 30~cm long and 6~cm in
diameter.
The beam was an 18.3~GeV/$c$ negatively charged 
beam consisting mainly (95\%) of $\pi^-$ and had a momentum bite $\Delta p/p$ of
3\% and a momentum resolution $\delta p/p$ of 1\%.
 The target to LGD distance was approximately 5~m.
The interacting beam trigger required no signal from a small scintillation counter
placed in the path of the beam  in coincidence
with beam defining counters upstream of the MPS.  Event triggers included all-neutral,
all charged and a mix of neutral plus charged particle topologies.  The neutral
particle requirement depended on energy and/or multi-photon effective mass information
from the LGD.

\subsection{\label{sec:data}Event reconstruction and selection}

We refer to the $ \pi^-\pi^0\pi^0p$ system as the \emph{neutral}
mode and the $\pi^-\pi^-\pi^+p$ system  as the \emph{charged} mode.  
In Tables~\ref{cutsn} and \ref{cutsc} the selection requirements and
resulting  event sample sizes are summarized for the neutral and charged
modes respectively.  In what follows we discuss details of these
requirements \cite{3piE852}.

\begin{table}
\caption{\label{cutsn}Summary of requirements  applied to the
data set leading to $\pi^-p \to \pi^-\pi^0\pi^0p$.  }
\begin{ruledtabular}
\begin{tabular}{lr}
Cut requirement & Number of Events\\
\hline
Satisfy topology trigger\footnotemark[1] & 123,748,800\\
Require four photons\footnotemark[1] & 13,737,265\\
Confidence level $> 0.20$\footnotemark[1] & 9,813,524\\
\hline
Beam hole\footnotemark[2] & 9,406,003\\
Vertex location\footnotemark[2] & 8,325,895\\
Recoil proton angle and momentum\footnotemark[2] & 6,620,596\\
CSI and DEA maximum energies\footnotemark[2] & 4,939,661\\
Photon separation\footnotemark[2] & 9,579,426\\
\hline
Final sample (all requirements) & 3,025,981\\
\end{tabular}
\end{ruledtabular}
\footnotetext[1]{Initial cuts, applied in the order shown.}
\footnotetext[2]{These cuts applied separately, after initial cuts.}
\end{table}

\begin{table}
\caption{\label{cutsc}Summary of requirements  applied to the
data set leading to $\pi^-p \to \pi^-\pi^-\pi^+p$.  }
\begin{ruledtabular}
\begin{tabular}{lr}
Cut requirement & Number of Events\\
\hline
Satisfy topology trigger\footnotemark[1] & 78,659,511\\
No photons\footnotemark[1] & 16,796,457\\
Confidence level $> 0.20$\footnotemark[1] & 10,157,455\\
\hline
Beam hole\footnotemark[2] & 8,038,452\\
Vertex location\footnotemark[2] & 8,038,452\\
Recoil proton angle and momentum\footnotemark[2] & 7,030,216\\
CSI and DEA maximum energies\footnotemark[2] & 4,822,330\\
$\Delta^{++}$ elimination\footnotemark[2] & 9,749,031\\
\hline
Final sample (all requirements) & 2,585,776\\
\end{tabular}
\end{ruledtabular}
\footnotetext[1]{Initial cuts, applied in the order shown.}
\footnotetext[2]{These cuts applied separately, after initial cuts.}
\end{table}

\subsubsection{Initial cuts}

The initial requirement was satisfaction of the \emph{topology} trigger.
For both modes a recoil charged track was required through
the tracking chamber (TCYL) surrounding the target.  The
neutral mode required a single forward-going charged particle 
in the PWCs along with energy deposition in the LGD corresponding
to multi-photon effective mass greater than 200~MeV/$c^2$.  For
the charged mode,  three
forward-going charged particles were required
in the PWCs.

\begin{figure}  
\centerline{\epsfig{file= 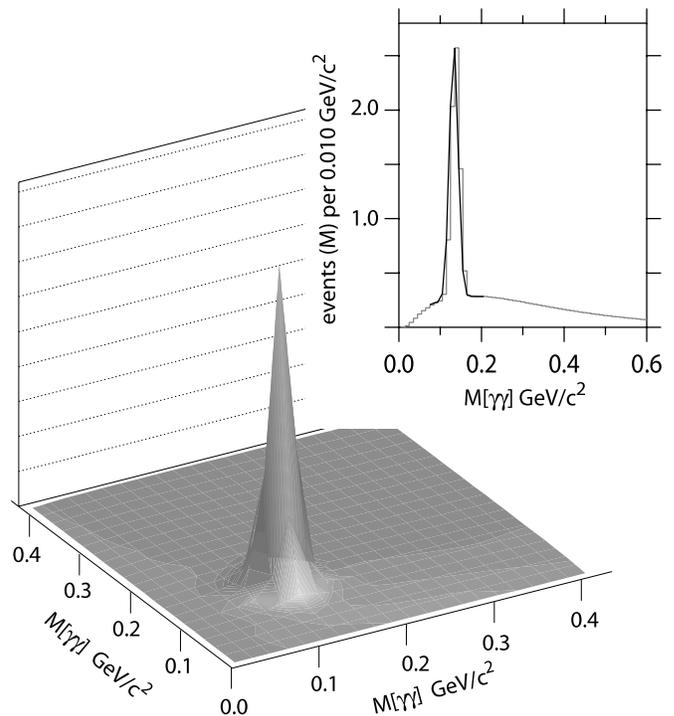,width= \columnwidth}}
\caption{
Scatterplot of one di-photon mass
against the other di-photon mas (three such pairings per event) for
events satisfying the $\pi^-\pi^0\pi^0p$ final state trigger and with
four reconstructed photons in the LGD.  The inset shows the 
di-photon mass spectrum with a fit to a Gaussian plus a polynomial.
The $\pi^0$ mass resolution is 10.4~MeV/$c^2$.
}\label{4gam}
\end{figure}

Four photons reconstructed in the LGD was the next requirement for the 
neutral mode.
The scatterplot of one di-photon mass
against the other di-photon mass is plotted in Figure~\ref{4gam} -- there are
three such pairings per event in this plot.  A clear peak corresponding to 
$\pi^0\pi^0$ production is observed. The inset of Figure~\ref{4gam}
shows a fit of the $2\gamma$ mass spectrum to a Gaussian and a polynomial.
The $\pi^0$ mass resolution is 10.4~MeV/$c^2$.
 For the charged mode further cuts
required no photons in the event.

\begin{figure}  
\centerline{\epsfig{file= 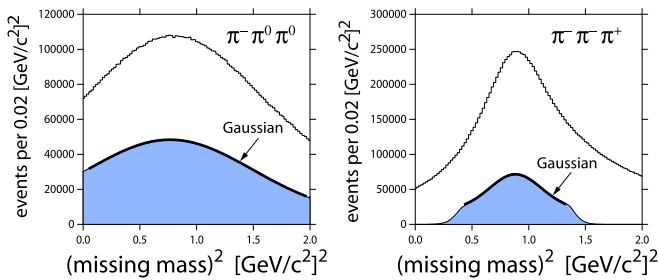,width=\columnwidth}}
\caption{
Distribution in missing mass squared recoiling against the $3\pi$
system for the (left)  $\pi^-\pi^0\pi^0$ mode and the
(right) $\pi^-\pi^-\pi^+$ mode.  The unshaded distribution is
before any cuts.  The shaded distribution is for events
that survived the confidence level cuts and all the cuts listed
in Tables~\ref{cutsn} and \ref{cutsc}.  The missing mass is
calculated before the kinematic fitting.  Results of fits of the
shaded distributions to a Gaussian are shown.
}\label{missmass}
\end{figure}

The distributions in missing mass squared recoiling against the 
$3\pi$ system for the neutral and charged modes are shown in
Figure~\ref{missmass} (unshaded). 
 Although a charged track corresponding
to the recoil proton is required, its momentum  is not
measured. A kinematic fitting program, SQUAW \cite{SQUAW}, is used to vary
the measured momenta of the the three pions and incoming beam,
within errors, to constrain the missing mass to be $m_p^2$.   For
the neutral mode, the two di-photon effective masses were 
constrained to the $\pi^0$ mass.

 \begin{figure}  
\centerline{\epsfig{file= 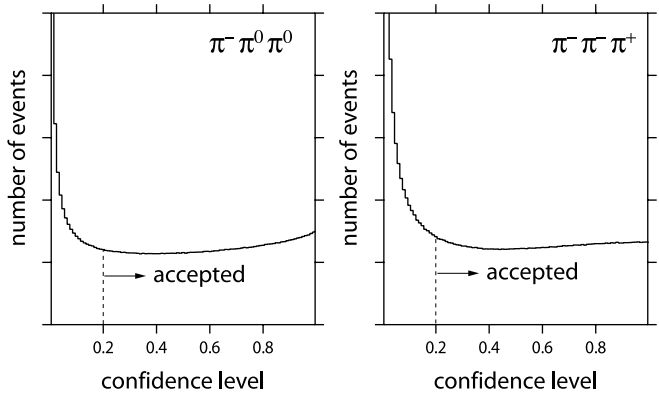,width=\columnwidth}}
\caption{
 Distribution in confidence level after kinematic
 fitting for the (left)  $\pi^-\pi^0\pi^0$ mode and the
(right) $\pi^-\pi^-\pi^+$ mode.
Events were required to have a confidence level greater
than 0.20.
}\label{cl}
\end{figure}

 After
the kinematic fit was applied the distribution in the confidence level ($C.L.$)
of the fit was observed to be flat for $C.L.>0.2$ (see Figure~\ref{cl}),
 and this cut was applied
to both neutral and charged modes.

\subsubsection{Other requirements}

The PWCs used to define the trigger topology were inefficient in the
region where the beam intercepted the chambers.  In order to 
eliminate uncertainties in the overall acceptance,  events
were rejected if a charged particle was within a $2.5 \ \sigma$ region
around the beam centroid at the plane of each chamber.  The
beam centroid and widths were determined from the data using
measurements from beamline elements.
Clearly, the fractional loss is
greater for the charged mode, with three  tracks,  than
the neutral mode with a single track.

The vertex of the reaction
was  required to be well-contained
within the target.  The $z$-position (along the beam direction) of the vertex
was required to be within the central 28~cm of the 30-cm~long target. 
In addition, the position of the vertex in the transverse
plane was required to be within 
2.5~$\sigma$ of the nominal beam position, resulting in an elliptical cut.  
This vertex cut, along with the confidence level cut, eliminated events with
$K_S \to \pi^+\pi^-$ or $K_S \to \pi^0\pi^0$.

The  momentum and
direction of the recoil proton can be inferred from the results of the
kinematic fit.  This inferred recoil direction was required to
be within 20$^{\circ}$ of the direction measured by the TCYL.  Furthermore
the computed magnitude of momentum of the recoil proton track had to be
consistent with the recoil proton having sufficient 
momentum to escape the LH$_2$ target.  The distance the recoil
proton travelled through the target was estimated based on the vertex
information and the angle of the proton as measured by TCYL.

The resolution of the charged particle tracking and  the
LGD were not sufficient to rule out events with low energy $\pi^{0\prime}$s
whose decay photons did not enter the LGD.  Thus the CSI and DEA were
used to eliminate such events based on energy deposition in these 
calorimeters.  Information from the TCYL was used to track charged particles
into segments of the CSI detector.  All segments without
a charged particle passing through them were required to register
less than 20~MeV of deposited energy.  Also, if no charged particle 
passed through the CPV counter in front of the DEA, the DEA was
required to register no signal.

\begin{figure}  
\centerline{\epsfig{file= 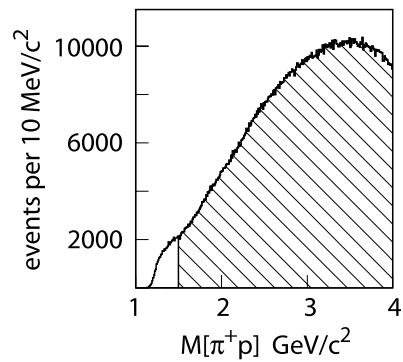,width=0.6\columnwidth}}
\caption{
Distribution in the $\pi^+p$ effective mass for events in the charged mode
satisfying the initial cuts.  The shaded distribution shows the events accepted
by the $m_{\pi^+p}>1.5$~GeV/$c^2$ requirement.
}\label{mpipr}
\end{figure}

The event selection requirements associated with the beam hole,
vertex location, recoil proton and energy deposition in the CSI and DEA
were applied identically to the neutral and charged modes.  
The remaining cuts applied were particular to the two different modes.
For the neutral mode it was required that the minimum photon cluster separation
was greater than 8~cm to insure reliable photon reconstruction.
 An additional cut for the charged mode required
 the $\pi^+p$ effective mass be greater than 1.5~GeV/$c^2$  to
eliminate a small signal due to $\Delta^{++}(1236)\to \pi^+p$ as shown in 
Figure~\ref{mpipr}.

After all cuts are imposed  3.0M~events remain  in the neutral mode
and 2.6M~events remain in the charged mode.  The 
resulting missing mass squared distribution
(before the kinematic fit) is shown in the shaded distributions of Figure~\ref{missmass}.

\subsection{\label{sec:masses}Effective mass and $t$ distributions}

\subsubsection{Uncorrected distributions}

\begin{figure}  
\centerline{\epsfig{file= 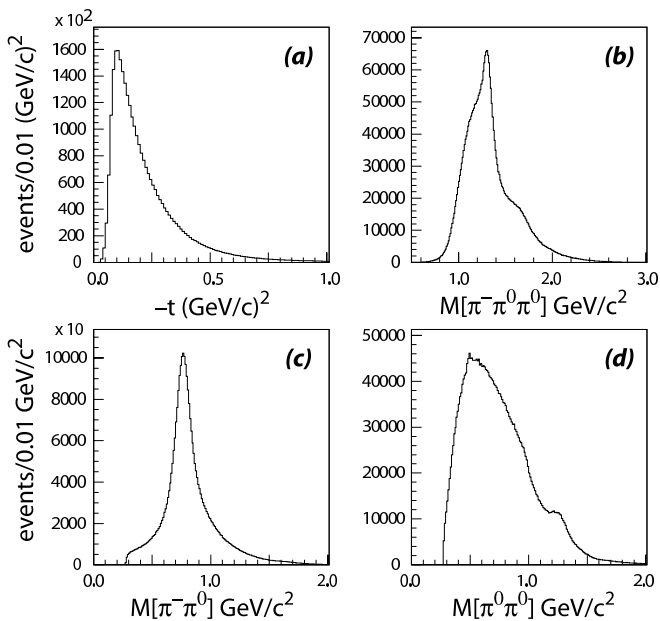,width=\columnwidth}}
\caption{
(a) Uncorrected
distribution in the square of the four-momentum transfer, $t$, from the
incoming $\pi^-$ to the outgoing $(3\pi)^-$;
(b) $(3\pi)^-$ effective mass distribution;
(c) $\pi^-\pi^0$ effective mass distribution; and
(d) $\pi^0\pi^0$ effective mass distribution for the neutral mode.
}\label{plots1_n}
\end{figure}

\begin{figure}  
\centerline{\epsfig{file= 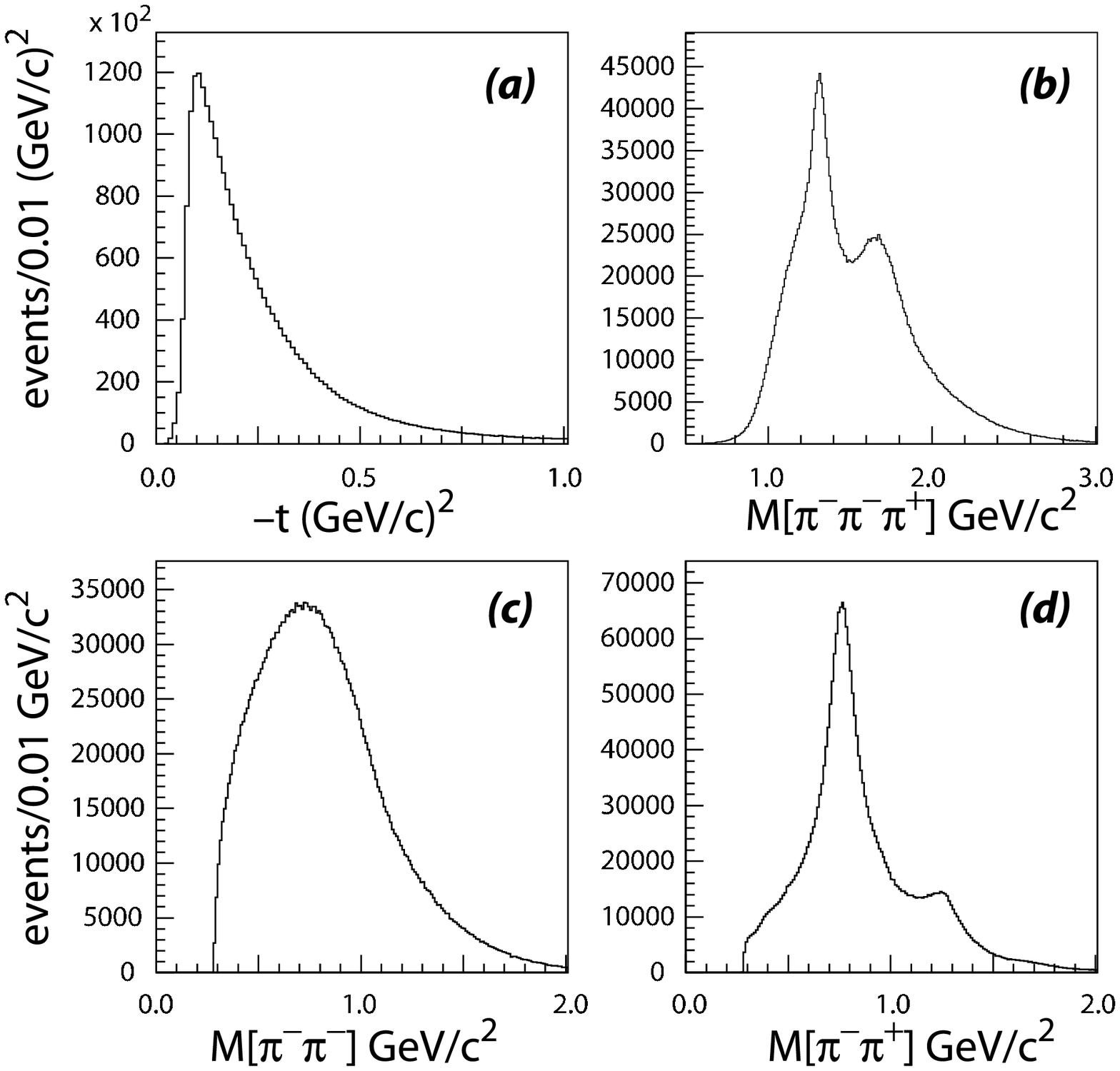,width=\columnwidth}}
\caption{
(a) Uncorrected
distribution in the square of the four-momentum transfer, $t$, from the
incoming $\pi^-$ to the outgoing $(3\pi)^-$;
(b) $(3\pi)^-$ effective mass distribution;
(c) $\pi^-\pi^-$ effective mass distribution; and
(d) $\pi^-\pi^+$ effective mass distribution for the charged mode.
}\label{plots1_c}
\end{figure}

Distributions of the
square of the four-momentum transfer, $t$,  from the incoming $\pi^-$ to the
outgoing $(3\pi)^-$ system as well as the $3\pi$ and $2\pi$ effective
mass distributions are presented in Figures~\ref{plots1_n}
and \ref{plots1_c}
 for the events passing the cuts listed in Tables~\ref{cutsn}
and \ref{cutsc}.  These distributions have not been corrected
for acceptance.
The distribution in $t$ for the neutral mode is shown in Figure~\ref{plots1_n}(a).
The partial wave analysis, discussed below, was carried out  for 12~bins in
$t$ defined in Table~\ref{tbins}.
The $\pi^-\pi^0\pi^0$ effective mass distribution is shown in Figure~\ref{plots1_n}(b).
A peak at the $a_2(1320)$ is observed as well as an enhancement in the
vicinity of the $\pi_2(1670)$.  The $\pi^-\pi^0$ mass distribution in Figure~\ref{plots1_n}(c)
shows clear indication of the $\rho(770)$ and the $\pi^0\pi^0$
in Figure~\ref{plots1_n}(d) shows the
$f_2(1275)$.

The distribution in $t$ for the charged mode is shown in Figure~\ref{plots1_c}(a).
The $\pi^-\pi^-\pi^+$ effective mass distribution is shown in Figure~\ref{plots1_c}(b).
Peaks at the $a_2(1320)$ and $\pi_2(1670)$
are clearly observed.  The $\pi^-\pi^-$ mass distribution in Figure~\ref{plots1_c}(c)
is featureless  and the $\pi^-\pi^+$
in Figure~\ref{plots1_c}(d) shows the $\rho(770)$ and the $f_2(1275)$.

\begin{table}
\caption{\label{tbins}Definition of the bins in $t$ -- 
square of the four-momentum transferred  from the incoming
$\pi^-$ to the outgoing $(3\pi)^-$ system -- used in this analysis.  }
\begin{ruledtabular}
\begin{tabular}{lr}
$t$-Bin & Range in (GeV/$c$)$^2$\\
\hline
$t1$ & 0.08 to 0.10\\
$t2$ & 0.10 to 0.12\\
$t3$ & 0.12 to 0.14\\
$t4$ & 0.14 to 0.16\\
$t5$ & 0.16 to 0.18\\
$t6$\footnotemark[1] & 0.18 to 0.23\\
$t7$ & 0.23 to 0.28\\
$t8$ & 0.28 to 0.33\\
$t9$ & 0.33 to 0.38\\
$t10$ & 0.38 to 0.43\\
$t11$ & 0.43 to 0.48\\
$t12$ & 0.48 to 0.53\\
\end{tabular}
\end{ruledtabular}
\footnotetext[1]{The results presented in this paper are for this $t$ bin unless noted
otherwise.}
\end{table}

\begin{figure}  
\centerline{\epsfig{file= 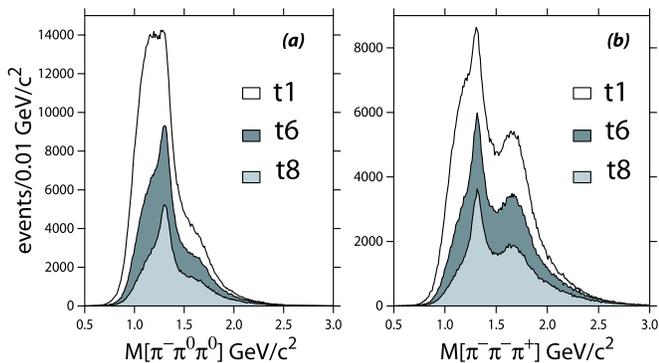,width=\columnwidth}}
\caption{Uncorrected
$(3\pi)^-$ mass distribution for the (a)~neutral mode and (b)~charged
mode for three $t$ regions:  $t_1$, $t_6$ and $t_8$ as defined in Table~\ref{tbins}.
}\label{mfort}
\end{figure}

Figure~\ref{mfort}(a) shows the  $\pi^-\pi^0\pi^0$ effective mass distribution 
for three regions of $t$ that we label $t1$, $t6$ and $t8$ and define in
Table~\ref{tbins}.
Figure~\ref{mfort}(b) shows the  $\pi^-\pi^-\pi^+$ effective mass distribution 
for the same three regions of $t$.  Clearly the overall shape of the 
$3\pi$ effective mass spectra
strongly depends on $t$.  These mass spectra are not corrected for
acceptance.

\begin{figure}  
\centerline{\epsfig{file= 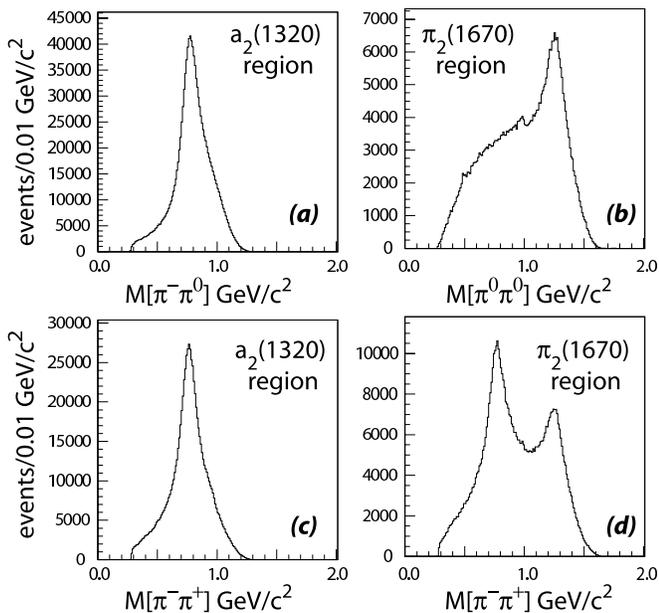,width=\columnwidth}}
\caption{Uncorrected 
$2\pi$ mass distributions for selected regions of the
$3\pi$ effective mass.  (a)~$\pi^-\pi^0$ mass in the $a_2$ region;
(b)~$\pi^0\pi^0$ mass in the $\pi_2$ region;
(c)~$\pi^-\pi^+$ mass in the $a_2$ region; and 
(d)~$\pi^-\pi^+$ mass in the $\pi_2$ region.  
}\label{2pi}
\end{figure}

The various $2\pi$ effective mass spectra also depend on $3\pi$ mass.
In Figure~\ref{2pi} we show several $2\pi$ spectra with the  requirement
that the $3\pi$ mass lie in the $a_2(1320)$ mass region
($1.20 < m_{3\pi} < 1.44$~GeV/$c^2$) or in the $\pi_2(1670)$ region
($1.54 < m_{3\pi} < 1.80$~GeV/$c^2$).  The $\pi^-\pi^0$ mass spectrum
(Figure~\ref{2pi}(a)) and $\pi^-\pi^+$ mass spectrum (Figure~\ref{2pi}(c))
are both dominated by the $\rho(770)$ when the $3\pi$ mass is in the
$a_2$ region.  The 
$\pi^0\pi^0$ mass spectrum
(Figure~\ref{2pi}(b)) and $\pi^-\pi^+$ mass spectrum (Figure~\ref{2pi}(d))
 both show strong signals for the  $f_2(1270)$ when the $3\pi$ mass is in the
$\pi_2$ region with the $\pi^-\pi^+$ mass spectrum also showing a
$\rho(770)$ resonance signal.

\subsubsection{Mass and $t$ acceptance and resolution}

The dependence of the  experimental acceptance  on 
 relevant kinematic variables was estimated by generating 
 Monte Carlo events for the two modes.  Events were
 generated with an exponentially damped distribution in $t$,
 based on what is observed in the data.
 The generated distribution in $3\pi$ mass was uniform from
 $3m_{\pi}$ to 2.5~GeV/$c^2$.  The $2\pi$ masses were 
 chosen to uniformly populate the $3\pi$ Dalitz plot at fixed
 $3\pi$ mass and the distributions in the relevant decay angles
 were also generated uniformly.
 
 The response of each detector component to a charged pion
 or photon from $\pi^0$ decay was simulated along with relevant
 smearing of momenta or energies.  The detector responses 
 were used as input to the same software used to reconstruct
 tracks and photons from actual data.  The output was then 
 passed on to the kinematic fitting software.  The cuts summarized
 in Tables~\ref{cutsn} and \ref{cutsc}  were then imposed on the
 simulated event samples.  
 
\begin{figure}  
\centerline{\epsfig{file= 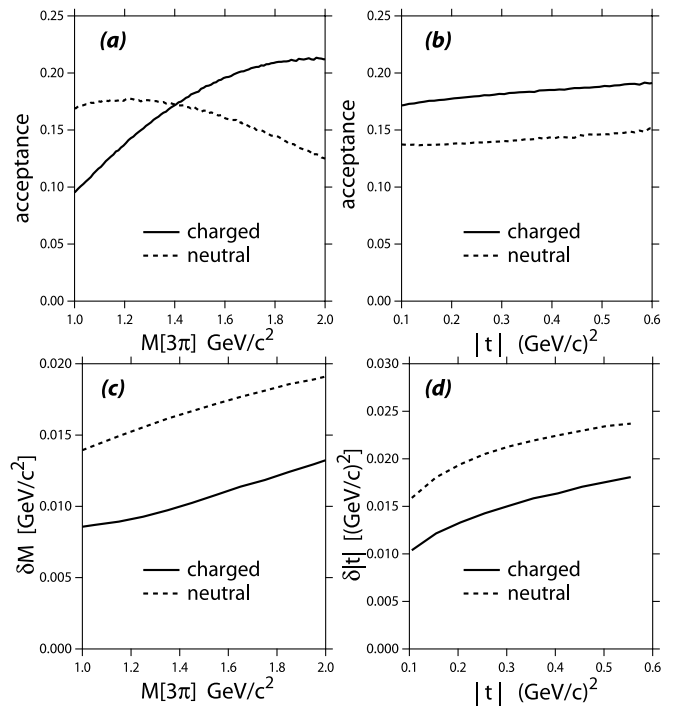,width=\columnwidth}}
\caption{
(a)~$3\pi$ mass acceptance as a function of $3\pi$ mass;
(b)~$t$ acceptance as a function of $t$;
(c)~$\delta m$ -- mass resolution as a function of $3\pi$ mass; and
(c)~$\delta t$ -- $t$ resolution as a function of $t$ mass -- all
for both neutral and charged modes.
}\label{m_and_t}
\end{figure}

The acceptance as a function of $3\pi$ effective mass, 
 for both neutral and charged modes, is shown
in Figure~\ref{m_and_t}(a).  The acceptance as a function of $t$ 
for both modes is shown in Figure~\ref{m_and_t}(b). 
The mass resolution, $\delta m$, as a function of  $3\pi$ effective mass
 and $t$ resolution, $\delta t$, as a function of $t$ for both 
 modes is shown in Figure~\ref{m_and_t}(c) and \ref{m_and_t}(d)
 respectively.
 
\begin{figure}  
\centerline{\epsfig{file= 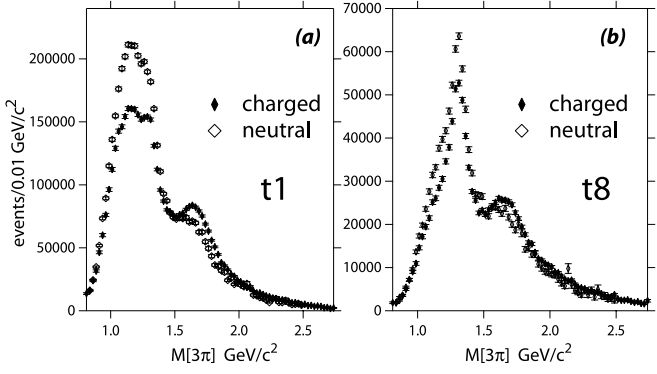,width=\columnwidth}}
\caption{
Acceptance corrected  $\pi^-\pi^0\pi^0$ and  $\pi^-\pi^-\pi^+$ 
effective mass distributions for two regions in $t$:
(a)~$t1$ and (b)~$t8$.  These $t$ regions are defined
in Table~\ref{tbins}.
}\label{macc_2t}
\end{figure}

Figures~\ref{macc_2t}(a) and \ref{macc_2t}(b) show the comparison 
between the acceptance-corrected $\pi^-\pi^0\pi^0$ and  $\pi^-\pi^-\pi^+$
mass distributions for two regions in $t$, $t1$ and $t8$ (defined
in Table~\ref{tbins}).  

\section{\label{sec:pwamethod}PWA methodology}

\subsection{\label{sec:pwaoverview}Overview}

\begin{figure}  
\centerline{\epsfig{file= 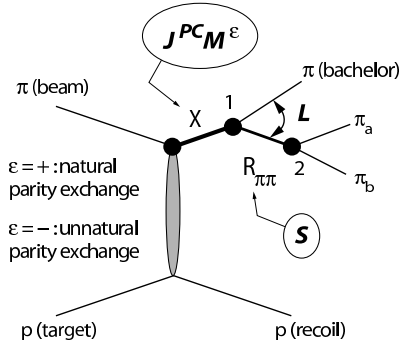,width=0.6\columnwidth}}
\caption{
Partial wave analysis of the $3\pi$ system in the isobar model.
The state $X$ is characterized by spin ($J$), parity ($P$) and
charge conjugation ($C$).  It decays at point~1 into a di-pion
resonance $R_{\pi\pi}$ (the isobar)
 and a bachelor $\pi$.  The di-pion has
spin $S$.  The angular momentum between the isobar 
and the bachelor $\pi$ is $L$.  At point~2 the di-pion resonance
decays into $\pi_a$ and $\pi_b$.  More details are given in the text.
}\label{pwa}
\end{figure}

The production of the $(3\pi)^-$ system in the reaction 
$\pi^-p \to (3\pi)^-p$ is described as a coherent and incoherent
sum of partial wave amplitudes. The production through one such partial wave
amplitude,
assuming the isobar model,
 is shown schematically in  Figure~\ref{pwa}.  
The state $X$
described by the partial wave
decays at point~1 into a di-pion resonance $R_{\pi\pi}$
(also referred to as the \emph{isobar})  and a bachelor $\pi$
followed by the decay of the di-pion resonance at point~2 into
$\pi_a$ and $ \pi_b$.  The state $X$ is characterized by 
$J^{PC}M^{\varepsilon}(SL)$, where $J$, $P$ and $C$ are the spin, parity and charge
 conjugation of the $3\pi$ system, respectively, $M$ is the spin projection along the $z$ axis  
and $\varepsilon$ represents symmetry (reflectivity)
of the $3\pi$ system under reflection in the 
production plane.  Viewed in terms of some exchange mechanism
(as in Figure~\ref{pwa}) $\varepsilon = +$ corresponds to natural
parity exchange and   $\varepsilon = -$ corresponds to unnatural
parity exchange.

 The decay at
point~1 is described in the Gottfried-Jackson
 frame \cite{GJ} which is the rest frame of
the $3\pi$ system with the beam direction defining the $z$-axis and the
normal to the production plane (specified by the momentum vectors of the
beam and recoil proton) defining the $y$-axis.  
In this frame the angles of the bachelor pion are denoted by
$\theta_{GJ}$ and $\phi_{GJ}$.
The angular momentum between the bachelor pion and
$R_{\pi\pi}$ is $L$, and the spin of $R_{\pi\pi}$ is $S$.
The decay at point~2 ($R_{\pi\pi} \to \pi_a\pi_b$)
 is described in the helicity frame -- the rest frame of the $\pi_a\pi_b$ system
 with the boost direction to that frame defining the $z$-axis.  
 In this frame the angles of one of the decay pions is denoted by
 $\theta_H$ and $\phi_H$.
 The angular
 distributions for the isobar decay are given by the spin of 
 $R_{\pi\pi}$ and the line shape as a function of $m_{\pi\pi}$
by a  relativistic Breit-Wigner function with a Blatt-Weisskopf factor \cite{Chung02}
with resonance parameters given by the Particle Data Group \cite{PDG}.
In this analysis the $S$-wave $\pi\pi$ isobars are parameterized 
according to the prescription described in reference~\cite{Chung02}.

 \begin{figure}  
\centerline{\epsfig{file= 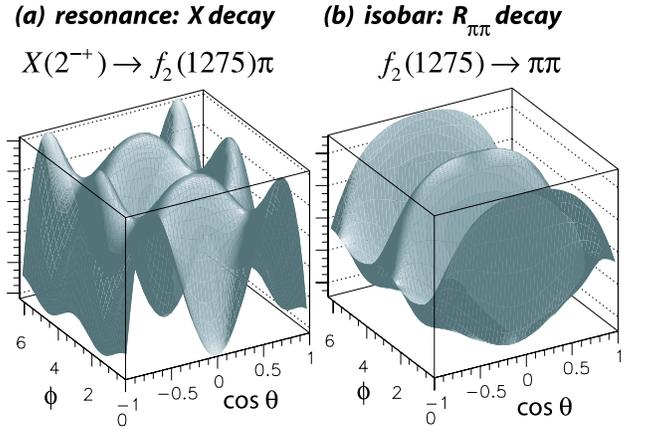,width=\columnwidth}}
\caption{
The decay angular structure for a particular partial wave where the
resonance ($X$) has $J^{PC}=2^{-+}$ and decays into $f_2(1275)\pi$
where the isobar is the $f_2(1275)$
and the angular momentum between the isobar and bachelor $\pi$
is $L=2$.
  (a)~The correlation between the
decay angles ($\theta_{GJ},\phi_{GJ}$) of the resonance in the Gottfried-Jackson
frame and (b) The correlation between the decay angles  ($\theta_{H},\phi_{H}$) 
of the isobar in the helicity frame.
}\label{sample}
\end{figure}

\begin{figure}  
\centerline{\epsfig{file= 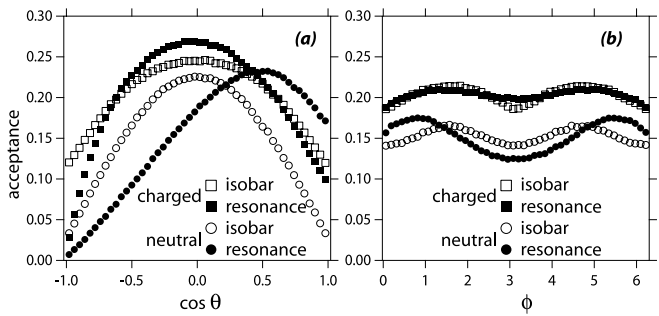,width=\columnwidth}}
\caption{
The acceptance in decay angles for the $3\pi$ effective mass in the 
$\pi_2(1670)$ mass region.  The angles $\theta$ and $\phi$ are measured in the
Gottfried-Jackson frame for the resonance decay and in the helicity
frame for the isobar decay.  Acceptances are shown
for (a) $\cos\theta$ and (b) $\phi$
 for both the charged
and neutral modes of the $3\pi$ system.
}\label{angles}
\end{figure}

Figure~\ref{sample} shows the correlated angular decay structure 
between $\theta_{GJ},\phi_{GJ}$
and $\theta_{H},\phi_{H}$ for one particular partial wave corresponding to the
decay of resonance $X$ with  $J^{PC}=2^{-+}$  into $f_2(1275)\pi$ where the isobar
is the  $f_2(1275)$
and the angular momentum between the isobar and bachelor $\pi$
is $L=2$.  A typical PWA involved fitting coherent or incoherent sums
of many such correlated distributions to the observed data, while taking into account
the acceptance of the experiment.  The acceptance in the relevant decay
angles for  the $3\pi$ effective mass in the 
$\pi_2(1670)$ mass region is shown in Figure~\ref{angles}.

The observed intensity as a function of $m_{3\pi}$ and $t$ is written as:

\begin{equation}
  I ( m_{3 \pi}, t, \tau ) =\eta(\tau) \sum_{\varepsilon} \left| \sum_b
  a_b^{\varepsilon} ( m_{3 \pi}, t ) A_b^{\varepsilon} ( \tau ) \right|^2
  \label{int}
\end{equation}
where $a_b^{\varepsilon} ( m_{3 \pi}, t )$
is the production amplitude,  $A_b^{\varepsilon} ( \tau ) $
is the decay amplitude and
 $\tau= \{ \theta_{GJ}, \phi_{GJ},  \theta_{H}, \phi_{H},m_{\pi\pi} \}$ is 
 the set of kinematic
  variables describing the 
decay of the resonance and isobar.
The index $b$ enumerates the spin variables of the partial
wave and $\varepsilon$ the reflectivity, as defined below.  The
$A_b^{\varepsilon} ( \tau ) $ involve products of $D^{J \star}_{M \lambda}$
functions for the two decays and a function describing the line
shape of the isobar.  In detail:

\begin{eqnarray}
    A_{JML}(\tau) =&&   \sqrt{(2L+1)(2S+1)}\nonumber\\
    &&\times
    \sum_{\lambda} D^{J \star}_{M
    \lambda} (\theta_{GJ},\phi_{GJ},0 ) D^{S \star}_{\lambda 0} 
    (\theta_{H},\phi_{H},0 )\nonumber\\
    &&\times
     \langle L 0 S \lambda |J \lambda \rangle\nonumber\\
    &&\times
    F_L ( p_X ) F_S ( p_R )  BW(m_{\pi\pi}) + (1 \to 2) \nonumber \\
\end{eqnarray}
where $1,2$ refer to the two identical pions ($\pi^-$'s in the charged mode and $\pi^0$'s in the neutral), 
 $J,M$ are the spin and projection of the resonance, 
$S$ is the spin of the isobar and  $L$ is the orbital angular momentum
between the isobar and the bachelor pion.  
The functions $F_L(p_X)$ and $F_S(p_R)$
 are the barrier factors as a function of the
 breakup momentum, $p_X$ and $p_R$,  for
 $X$ and the isobar respectively.
$BW(m_{\pi\pi})$
is the Breit Wigner function for the isobar. 
 In the reflectivity
basis:

\begin{equation}
    A_{JML}^{\varepsilon} =  \Theta(M)   
  \left[  A_{JML} - \varepsilon P (-1)^{J-M} A_{J-ML} 
     \right]
\end{equation}
where $P$ is the parity of the
resonance,  $\Theta(0) =1/2$, $\Theta(M>0)=\sqrt{1/2}$ and
$\Theta(M<0)=0$.

The goal of the PWA is to extract the production amplitudes by fitting the
correlated decay angular distributions to the intensity described by 
equation~\ref{int} in bins of $m_{3\pi}$ and $t$.  In order to account for
detector acceptance, Monte Carlo simulations were used to find the
acceptance function $\eta(\tau)$ in terms of the set of kinematic variables
$\tau$.  Thus the data were fitted to the decay amplitudes modified by
the acceptance function.

The PWA formalism used here is identical to that used in \cite{Chung02}
and \cite{Chung93} and is also briefly described in  internal
notes \cite{3piE852}.
The PWA software was developed
at Brookhaven Lab and  Indiana University (IU)
 \cite{3piE852}.    The IU software was optimized for running on a 200-processor
computer cluster (AVIDD) \cite{AVIDD} allowing systematic studies involving
 many PWA fits, varying isobar parameters
and using different wave sets. 
The IU programs were also used to 
analyze the data presented in \cite{Chung02}
and produced consistent results (see Appendix~\ref{sec:online}
for a list of technical notes available online).

\subsection{\label{sec:pwaselect}Selecting a partial wave set}

\subsubsection{General criteria}

As noted earlier, it is important to include in the analysis a set of partial waves 
sufficient to describe the physics and to have well understood criteria for
selecting this set.
 One technique,
following \cite{Daum81}, is 
to start with some parent set of waves and examining the change in
 likelihood resulting from sequentially removing waves.

  \begin{table}
\caption{\label{thewaves}Partial waves under consideration in this analysis. }
\begin{ruledtabular}
\begin{tabular}{llll}
$J^{PC}M^{\varepsilon}$ & $L$ & Isobar $\pi$ & Wave Set\footnotemark[1]\\
\hline
$  0^{-+}0^+$ & S & $(\pi \pi)_S \pi$ & B\\
$  0^{-+}0^+$ & S & $f_0\pi$ & B\\
$  0^{-+}0^+$ & P & $\rho\pi$ & B\\
\hline
$  1^{++}0^+$ & S & $\rho\pi$   & B\\
$  1^{++}1^+$ & S & $\rho\pi$   & B\\
$  1^{++}1^-$ & S & $\rho\pi$   & L\\
$  1^{++}0^+$ & P &  $f_0\pi$  & H\\
$  1^{++}0^+$ & P & $(\pi \pi)_S \pi$   & H\\
$  1^{++}1^+$ & P &  $f_2\pi$  & H\\
$  1^{++}0^+$ & D & $\rho\pi$   & L\\
\hline
$  1^{-+}1^+$ & P & $\rho\pi$   & B\\
$  1^{-+}0^-$ & P & $\rho\pi$   & B\\
$  1^{-+}1^-$ & P & $\rho\pi$   & B\\
\hline
$  2^{++}1^+$ & D & $\rho\pi$   & B\\
$  2^{++}0^-$ & D & $\rho\pi$   & B\\
\hline
$  2^{-+}0^+$ & S & $f_2\pi$   & B\\
$  2^{-+}1^+$ & S & $f_2\pi$   & B\\
$  2^{-+}1^-$ & S & $f_2\pi$   & L\\
$  2^{-+}0^+$ & P &  $\rho\pi$  & B\\
$  2^{-+}1^+$ & P &  $\rho\pi$  & H\\
$  2^{-+}0^+$ & P &  $\rho_3\pi$  & H\\
$  2^{-+}1^+$ & P &  $\rho_3\pi$  & H\\
$  2^{-+}0^+$ & D &  $(\pi \pi)_S \pi$ & B\\
$  2^{-+}0^+$ & D &  $f_0\pi$ & H\\
$  2^{-+}1^+$ & D &   $(\pi \pi)_S \pi$ & H\\
$  2^{-+}1^+$ & D &   $f_0\pi$ & H\\
$  2^{-+}1^+$ & D &   $f_2\pi$ & B\\
$  2^{-+}0^+$ & D &  $f_2\pi$ & B\\
$  2^{-+}0^+$ & F &  $\rho\pi$ & H\\
$  2^{-+}1^+$ & F &  $\rho\pi$ & H\\
\hline
$  3^{++}0^+$ & S & $\rho_3\pi$ & B\\
$  3^{++}0^+$ & P & $f_2\pi$ & H\\
$  3^{++}0^+$ & D & $\rho\pi$ & H\\
\hline
$  4^{++}0^+$ & D & $\rho_3\pi$ & H\\
$  4^{++}0^+$ & F &  $f_2\pi$ & H\\
$  4^{++}0^+$ & G & $\rho\pi$ & H\\
\hline
$  4^{-+}0^+$ & P & $\rho_3\pi$ & H\\
$  4^{-+}0^+$ & F & $\rho\pi$ & H\\
\hline
Background &  &  & B\\
\end{tabular}
\end{ruledtabular}
\footnotetext[1]{Indicates whether wave was used in the high wave
set alone [H], the low wave set alone [L] or in both sets [B].}
\end{table}

Another arbiter of wave set sufficiency is the comparison of moments, $H(\ell mm^{\prime})$, 
of the $D^{\ell}_{mm^{\prime}}(\Omega)$ functions
as calculated directly from the data and as calculated using the
results of PWA fits.  A PWA using a sufficient set of partial waves
will reproduce all angular distributions of the data as evidenced by
the agreement of the observed and calculated moments.
We define $H(\ell mm^{\prime}) =
 \int  I(\Omega)D^{\ell}_{mm^{\prime}}(\Omega)d\Omega$ where $\Omega$
 represents the Euler angles of the $3\pi$ system
 and the intensity $I(\Omega)$ is determined directly from experiment
 or computed using the results of the PWA fits.  Both of these techniques were 
 used in this analysis.

\subsubsection{Defining a parent set of waves}

The $\pi\pi$ isobars included in this analysis are the
 $ \sigma$, $f_0(980)$, $\rho(770)$,
 $f_2(1275)$ and $\rho_3(1690)$
 where $\sigma$ is meant to indicate a $S$-wave
 $\pi\pi$ system as described in \cite{Chung02}.  In
 addition, a background wave is included.  The background wave is characterized
by a uniform distribution in the relevant decay angles and is
added incoherently with the other waves.   We restrict the
parent wave set to those with $ J\leq 4$, $M \leq 1$ and $S \leq 3$.
 
 \subsubsection{Removing waves from the parent set}
 
\begin{figure}  
\centerline{\epsfig{file= 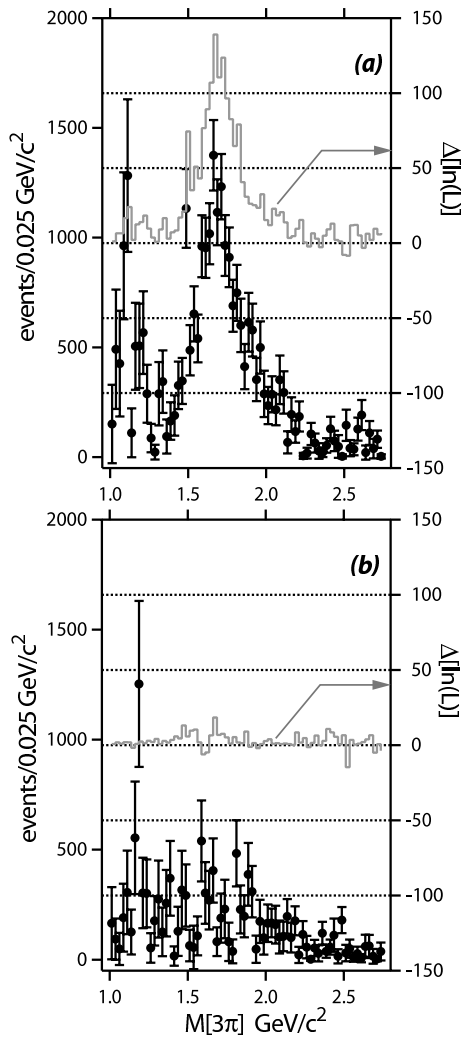,width=0.7\columnwidth}}
\caption{
Partial wave intensity (filled circles) as a function of $3\pi$ mass
(lower plot) shown along with the change in $ \ln \cal{L} $
(shown above the intensity plot -- read the right axis scale)
that results when this partial wave is removed from the parent
set of partial waves.  (a)~$2^{-+}1^{+}f_2\pi D\ $wave - a significant
wave and (b)~$2^{-+}1^{-}f_2\pi D\ $wave - an insignificant wave.
}\label{sig_nosig}
\end{figure}

Waves were sequentially removed from the parent set
and the resulting change
in likelihood ($\cal{L}$) examined  \cite{3piE852}.
Three passes were used.  First, waves with an L=0 isobar
($ \pi \pi $ $S$-wave and $f_0$) were examined. 
Subsequent passes examined $L=2$ ($f_2$) and 
$L=1$ ($ \rho$) isobar partial waves.
A PWA with the wave in question removed was performed and
the resulting likelihood compared to PWA including the wave.
If the change in $ \ln \cal{L} $ was less than 40 the wave was removed.
Figure~\ref{sig_nosig} shows the change in $ \ln \cal{L} $ 
when a significant, but not a dominant,  wave ($2^{-+}1^{+}f_2\pi D\ $wave) is
removed compared to when an insignificant  wave
($2^{-+}1^{-}f_2\pi \ D\ $wave) is removed.
 
Certain partial waves that would have been removed by this selection
criterion were kept because the existence of signals in these waves
is under consideration.
The negative reflectivity exotic $1^{-+}$ partial waves were kept even though they would have been
chosen for removal by the above criterion. 
The positive reflectivity exotic partial wave failed the selection criterion everywhere
above 1.4 ${\rm{GeV}}/c^2$ and exhibited large fluctuations in intensity below
this mass. 
This partial wave was also kept and its contribution to the likelihood is further
examined below using the final set of partial waves.
Finally, the $ 2^{++}0^{-}\rho \pi\ D$-wave (which failed the selection test) 
was kept to serve as a possible reference wave for evaluation of
phase differences of negative reflectivity partial waves.
Based on these criteria the compiled set 
includes 35 waves  and a background wave.  We refer to this as the \emph{high-wave set}.

The analysis reported in reference~\cite{Chung02} used a wave set consisting
of 20 waves including the background wave -- we refer to this as the \emph{low-wave set}.
 Table~\ref{thewaves} lists the waves used in low-wave and high-wave sets.
The waves included only in the low-wave set are labeled with [L] in
 Table~\ref{thewaves} while the 
 waves included only in the high-wave set are labeled with [H] and those
 included in both sets are labeled with [B]. 

\subsubsection{Using moments}

\begin{figure}  
\centerline{\epsfig{file= 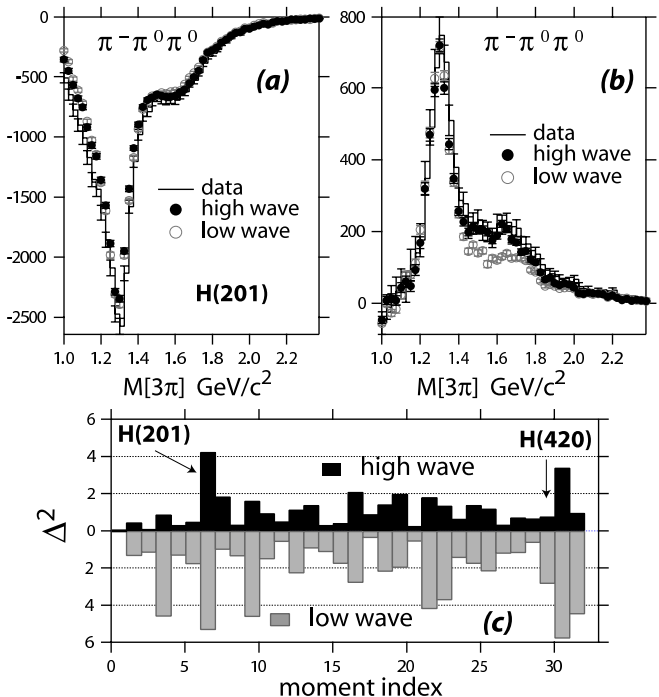,width=\columnwidth}}
\caption{
Comparison of the (a) H(201) and (b) H(420)
 moments as computed directly from data and 
from PWA fits for the low and high wave sets for the $\pi^-\pi^0\pi^0$
channel.  In (c) the $\Delta^2$ (differences squared
divided by errors squared summed over all mass bins
and divided by the number of mass bins) is shown for various
moments. 
}\label{fig2}
\end{figure}

\begin{figure}  
\centerline{\epsfig{file= 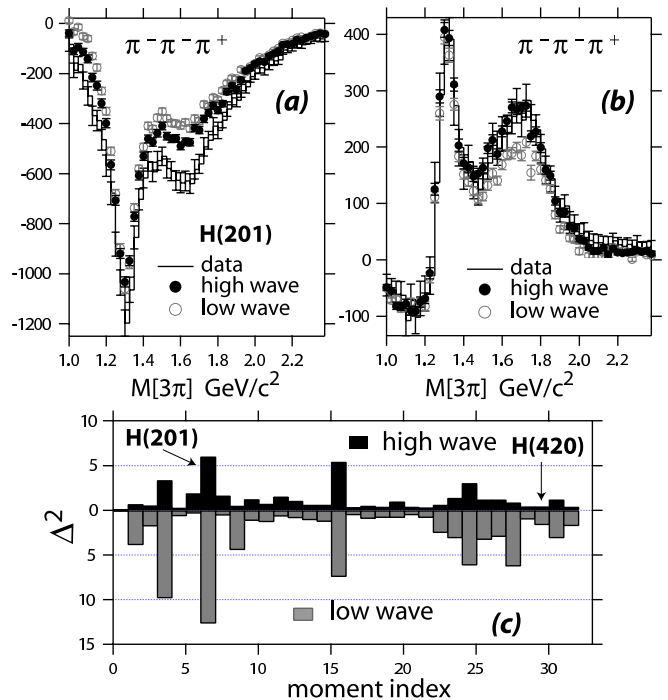,width=\columnwidth}}
\caption{
Comparison of the (a) H(201) and (b) H(420)
 moments as computed directly from data and 
from PWA fits for the low and high wave sets for the $\pi^-\pi^-\pi^+$
channel.  In (c) the $\Delta^2$ (differences squared
divided by errors squared summed over all mass bins
and divided by the number of mass bins) is shown for various
moments.  
}\label{fig3}
\end{figure}

The PWA solutions obtained with the two wave sets were used to compute
moments that were then compared directly with data.  
In Figure~\ref{fig2}
we show the $H(201)$ and $H(420)$ moment comparisons
as a function of $3\pi$ mass for the
$\pi^-\pi^0\pi^0$ mode for low-wave and high-wave set PWA fits.
We also show the difference between data and PWA calculations
of the moments as  $\Delta^2$ (summing differences squared
divided by errors squared summed over all mass bins
and divided by the number of mass bins) for various
moments.  
Similar plots are shown in Figure~\ref{fig3} for the $\pi^- \pi^-\pi^+ $ mode.
The moments calculated using the PWA results from the high wave set
have better agreement with experimental moments, but there are
moments (such as $H(201)$ - shown in part (a) of Figure~\ref{fig3})
for which agreement is not achieved.  This may be due to the inherent
inadequacy of the isobar model in describing the underlying production
mechanism.  

An examination of Figures~\ref{fig2}(c) and \ref{fig3}(c) show that
for all the moments,  the moments calculated using the high-wave set
have better agreement with data than do moments 
 calculated using the low-wave set.

\section{\label{sec:pwaresults}PWA results}
 
 The PWA results presented in this section
  were for data with $0.18 < |t| < 0.23$~(GeV/$c$)$^2$
  referred to as $t6$ in Table~\ref{tbins}.   The analysis was also carried out
in 11 other bins in $|t|$ -- some results of which will also be summarized below.

\begin{figure}  
\centerline{\epsfig{file= 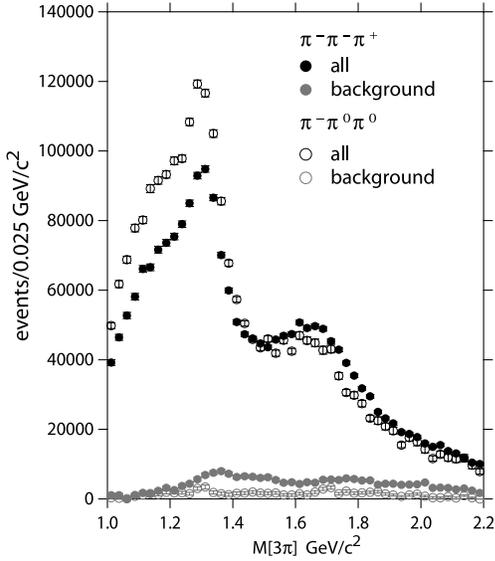,width=0.75\columnwidth}}
\caption{
The sum of a partial waves and the background wave for both the
charged and neutral modes.
}\label{nc_all_bkg}
\end{figure}

Figure~\ref{nc_all_bkg} shows the sum of all waves in the high-wave set
along with the background wave for both the charged and neutral modes.
The fitting procedure constrains the sum of all the waves to 
agree with the acceptance-corrected $3\pi$ mass distribution.  The
background wave (uniform in all angles) is added incoherently with
the other waves.  As can be seen, the contribution from the background wave
is small (of order 1\%).

There is a systematic disagreement in the relative normalization between
the two modes resulting from an overall per track inefficiency of 5\% in
the trigger PWC's.  The overall inefficiency has no bearing on the
features of the acceptances and was thus not incorporated into the Monte
Carlo corrections, but results in the neutral mode having a yield
consistently 25\%
higher than the charged mode.  The
neutral mode required only one forward-going charged 
particle in the trigger while the charged
mode required three.

In the following we discuss intensities and phases of 
individual partial waves. A partial wave intensity corresponds to a single term of 
equation~(\ref{int}) which is a product of the  fitted partial wave amplitude,
 $a^\epsilon_b(m_{3\pi},t)$,  and the isobar amplitude $A_{JML}(\tau)$ . 
Similar to the procedure followed in references~\cite{Adams98,Chung02}, 
 the  intensities are integrated over acceptance corrected data, and are
  normalized to the total number of events.

\begin{figure}  
\centerline{\epsfig{file= 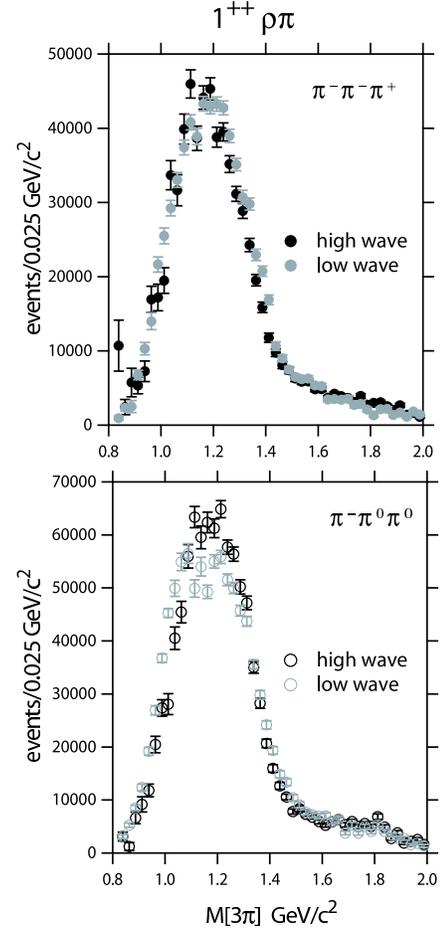,width=0.65\columnwidth}}
\caption{
Coherent sum of the $1^{++}$ $\rho\pi$-waves for the 
(upper)   $\pi^-\pi^-\pi^+$    and (lower)  $\pi^-\pi^0\pi^0$  modes.
For each the results from the high wave set and low wave set 
are shown.
}\label{coherent1pp}
\end{figure}

 The coherent sum of the $1^{++}$ $\rho\pi$-waves is shown in 
 Figure~\ref{coherent1pp} for the $\pi^-\pi^-\pi^+$ and $\pi^-\pi^0\pi^0$ modes.
 For each the results from the high wave set and low wave set 
are shown.  The  $1^{++}$ wave is a dominant wave and the results of
the PWA for the two wave sets (high and low) are in good agreement.  The
yield for the two charged modes is also consistent with expectations from
isospin for the $\rho\pi$ system.

The  $1^{++}$ $\rho\pi$ $D$-wave is not included in the high-wave set,
since it did not meet criteria for inclusion, but it is included in the low-wave
set.  The observed ratio of $1^{++}$ $\rho\pi$ $S$-wave to $D$-wave
in the low-wave set is consistent with that of reference~\cite{Chung02}.

\begin{figure}  
\centerline{\epsfig{file= 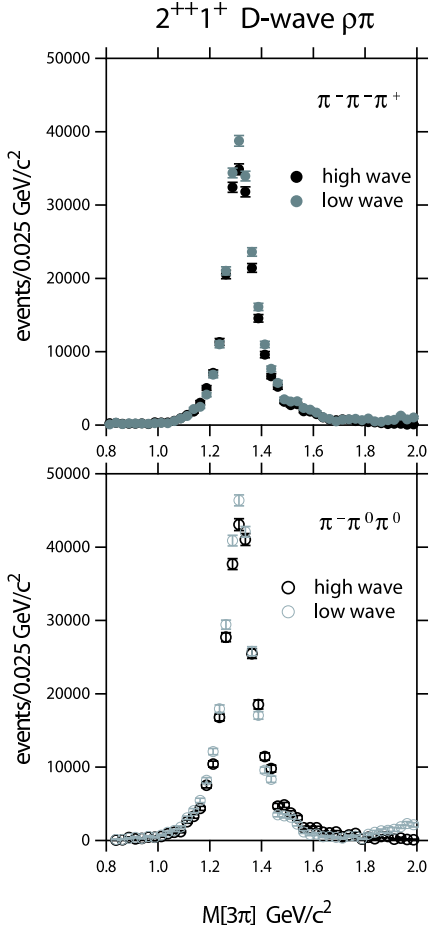,width=0.65\columnwidth}}
\caption{
$2^{++}1^+ \ D$-wave $\rho\pi$ partial wave intensity for the 
(upper) $\pi^-\pi^-\pi^+$ and (lower) $\pi^-\pi^0\pi^0$ modes.
For each the results from the high wave set and low wave set 
are shown.
}\label{2pp}
\end{figure}

 Figure~\ref{2pp} shows the $2^{++}1^+ \ D$-wave $\rho\pi$ partial wave as
 a function of $m_{3\pi}$ for both the neutral and charged modes. 
A clear $a_2(1320)$ signal  is observed.  For a pure $\rho\pi$
isovector resonance, we expect an equal number of charged mode
and neutral mode events.
 This wave
is clean and robust and will be used for interferometry to study the phase
motion of other waves relative to this wave.

\begin{figure}  
\centerline{\epsfig{file= 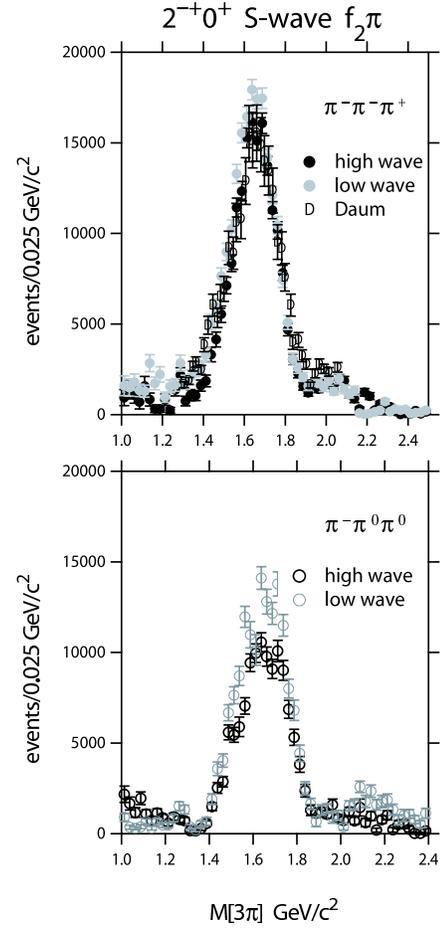,width=0.65\columnwidth}}
\caption{
$2^{-+}0^+ \ S$-wave $f_2\pi$ partial wave intensity for the 
(upper) $\pi^-\pi^-\pi^+$ and (lower) $\pi^-\pi^0\pi^0$ modes.
For each the results from the high wave set and low wave set 
are shown.
}\label{2mpS}
\end{figure}

\begin{figure}  
\centerline{\epsfig{file= 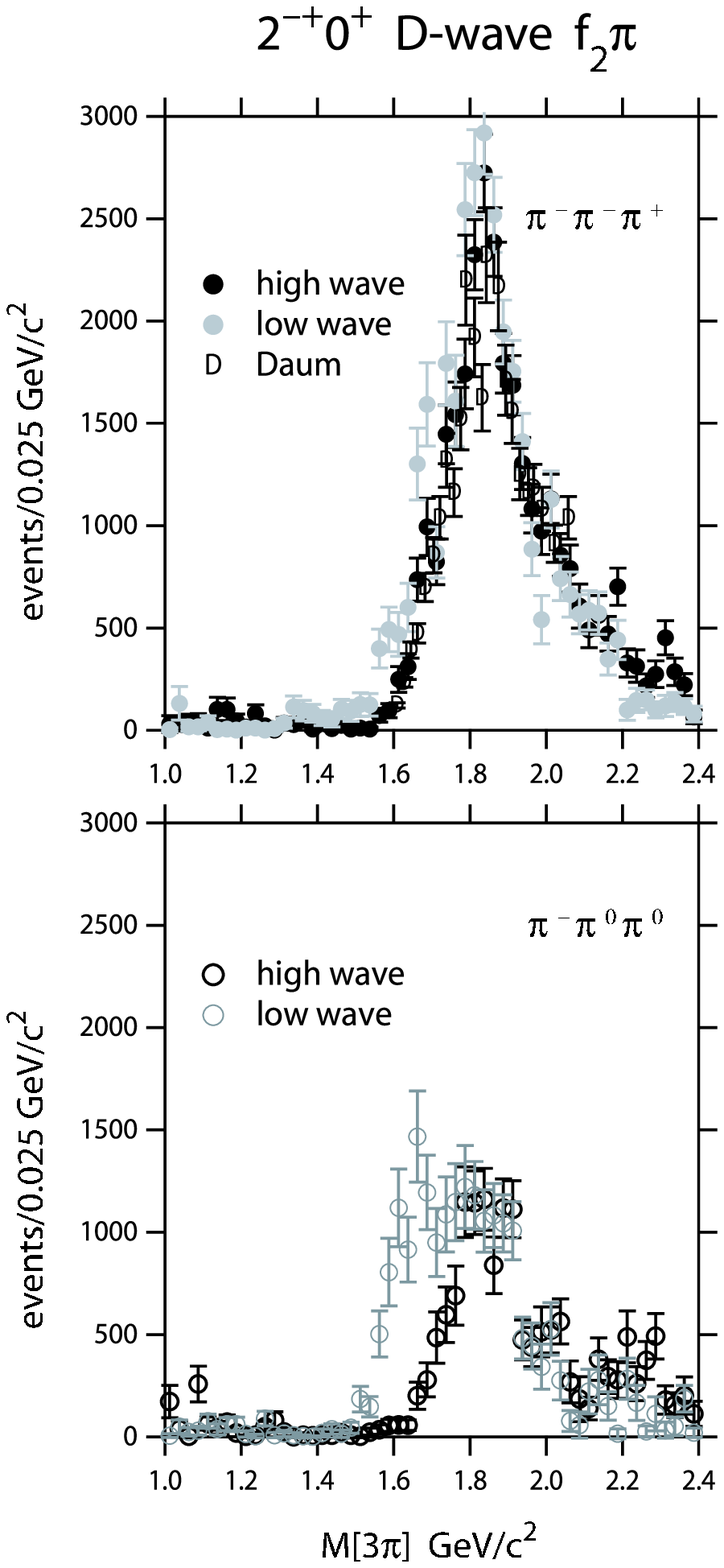,width=0.65\columnwidth}}
\caption{
$2^{-+}0^+ \ D$-wave $f_2\pi$ partial wave intensity for the 
(upper) $\pi^-\pi^-\pi^+$ and (lower) $\pi^-\pi^0\pi^0$ modes.
For each the results from the high wave set and low wave set 
are shown.
}\label{2mpD}
\end{figure}

 Figures~\ref{2mpS} and \ref{2mpD} show the intensities for the 
 $2^{-+}0^+ \ S$-wave $f_2\pi$ and $2^{-+}0^+ \ D$-wave $f_2\pi$
 waves respectively for both the charged and neutral modes.  
 Note that the relative yields between charged and neutral modes,
 for these $f_2\pi$ decays are consistent with expectations from isospin,
  after taking into account the overall 25\% reduction in yield for the
 charged mode.   These 
 intensities are also compared with those measured by
 Daum~{\it et. al.}  \cite{Daum81} in a similar amplitude analysis
 of diffractive $3\pi$ production in $\pi^-p$ interactions
 at 63 and 94~GeV/$c^2$.
The Daum results have been arbitrarily scaled by a factor of 5.

 The $S$ and $D$ $f_2\pi$ modes  both show a resonant-like structure but not at the same peak position. The $S$ wave peaks at  approximately 1.6~GeV/$c^2$ while the $D$ wave peaks at approximately 1.8~GeV/$c^2$. This shift in the peak position has been observed in other analyses~\cite{Chung02, Daum81}.  This
has been interpreted as
the  $f_2\pi$ decay mode of two different resonances. The PDG places the $\pi_2(1670)$ $2^{-+}$ resonance at a mass of 1.67~GeV/$c^2$ based on fits to the $S$ wave mode alone.

  \begin{figure}  
\centerline{\epsfig{file= 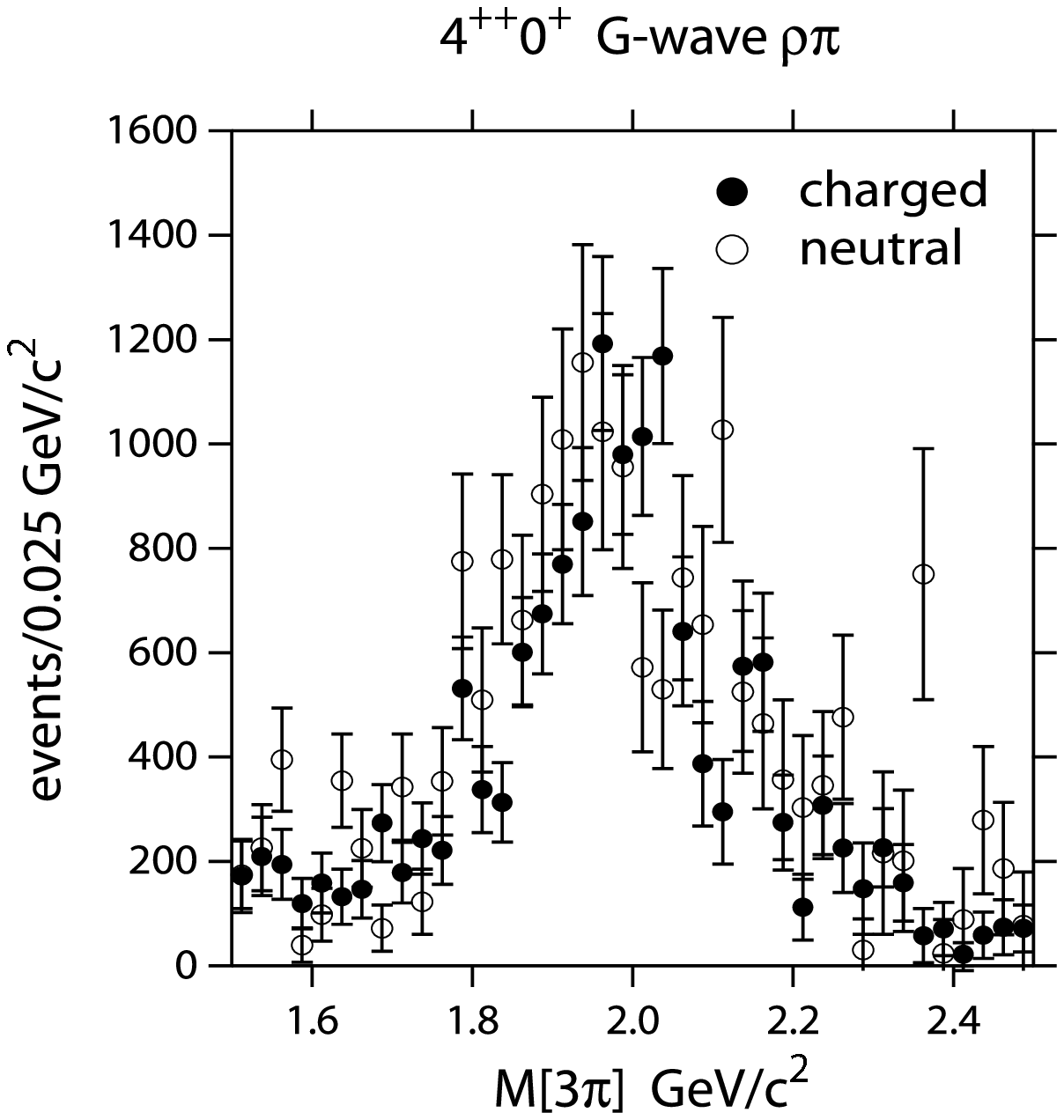,width=0.75\columnwidth}}
\caption{
The $4^{++}0^+ \ G$-wave $\rho\pi$ partial wave intensity 
for the $\pi^-\pi^-\pi^+$ and $\pi^-\pi^0\pi^0$ modes.
}\label{a4}
\end{figure}  
 
  Figure~\ref{a4} shows the $4^{++}0^+ \ G$-wave $\rho\pi$ intensity as
 a function of $m_{3\pi}$ for both the neutral and charged modes.  
Comparing Figures~\ref{a4} and \ref{2pp} the yield of $a_4(2040)$
is about 3\% that of the $a_2(1320)$ indicating the sensitivity
of the PWA within the isobar model for finding states with relatively
low production cross-sections.

\section{\label{sec:exotica}The exotic $1^{-+}$ wave}

\subsection{Comparing the high and low wave sets}

 \begin{figure}  
\centerline{\epsfig{file= 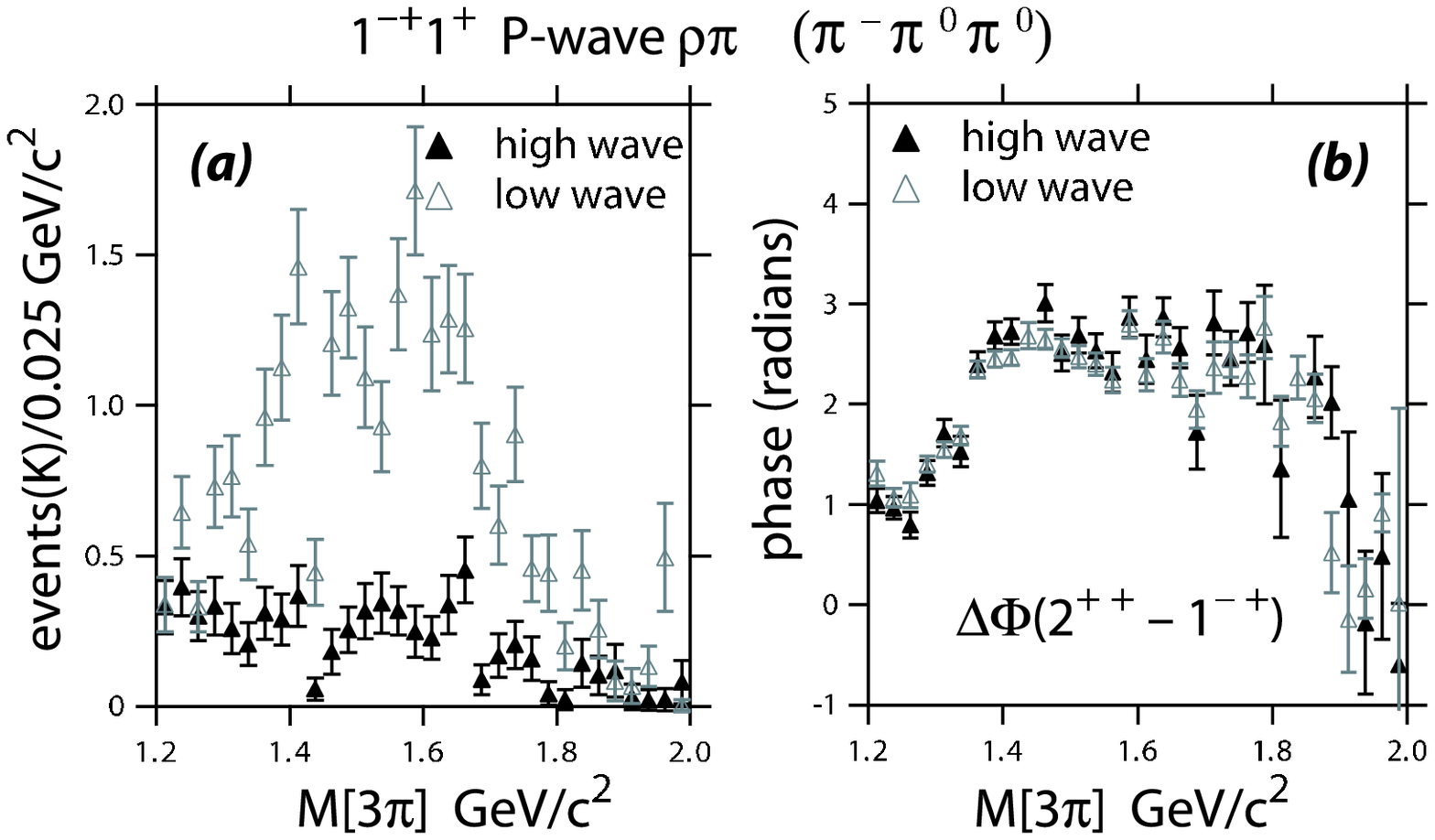,width=0.95\columnwidth}}
\caption{
(a)  The $1^{-+}1^+ \ P$-wave $\rho\pi$ partial wave in the neutral
mode ($\pi^-\pi^0\pi^0$) for the high-wave set PWA and the low-wave
set PWA and (b)~the phase difference $\Delta\Phi$ between the
$2^{++}$ and $1^{-+}$ for the two wave sets.
}\label{exotic_n}
\end{figure}

 \begin{figure}  
\centerline{\epsfig{file= 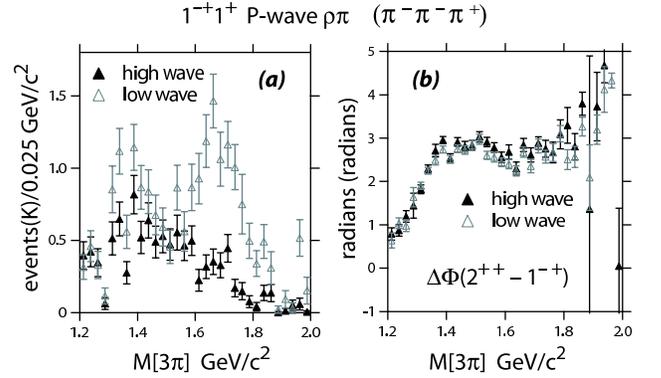,width=0.95\columnwidth}}
\caption{
(a)  The $1^{-+}1^+ \ P$-wave $\rho\pi$ partial wave in the charged
mode ($\pi^-\pi^-\pi^+$) for the high-wave set PWA and the low-wave
set PWA and (b)~the phase difference $\Delta\Phi$ between the
$2^{++}$ and $1^{-+}$ for the two wave sets.
}\label{exotic_c}
\end{figure}

The $1^{-+}1^+ \ P$-wave $\rho\pi$ partial wave  and its interference
with the $2^{++}$ wave are shown in
 parts (a) and (b) of Figures~\ref{exotic_n} and \ref{exotic_c}
 for the neutral mode ($\pi^-\pi^0\pi^0$)  and charged 
 mode ($\pi^-\pi^-\pi^+$).
 When
the low-wave set is used in the fit, an enhancement is observed
for the exotic $1^{-+}$ wave in the $3\pi$
mass region around 1.6~GeV/$c^2$, consistent with the observation of 
\cite{Adams98,Chung02}.
The enhancement disappears when the
high wave set is used.  This
effect is observed for both $3\pi$ channels. 
The phase  of the
exotic wave relative to the dominant $2^{++}$ wave
(see Figures~\ref{exotic_n}(b) and \ref{exotic_c}(b))
 is similar for
both wave sets and dominated by the phase of the $a_2(1320)$.
The apparent phase motion beginning around 1700~MeV/$c^2$ is likely due to
the $a_2(1700)$~\cite{PDG}, but the unreliability of the exotic intensity
makes this an inappropriate place for its study.

\subsection{Leakage of the $\pi_2(1670)$ and the instability of the exotic wave}

\begin{figure}  
\centerline{\epsfig{file= 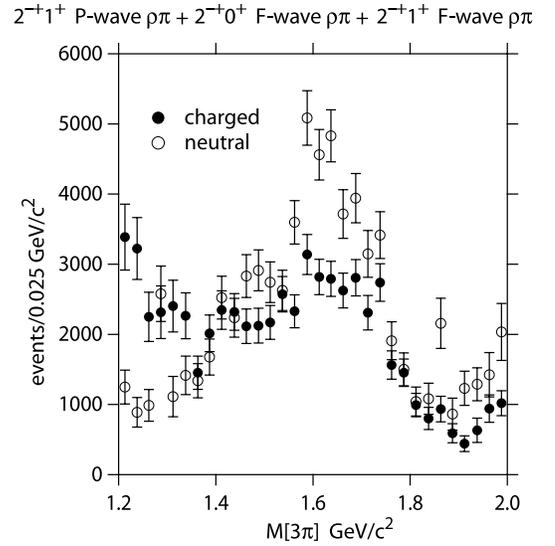,width=0.80\columnwidth}}
\caption{
Sum of the intensities for the $2^{-+}0^+ \ F$-wave $\rho\pi$,
$2^{-+}1^+ \ F$-wave $\rho\pi$ and $2^{-+}1^+ \ P$-wave $\rho\pi$
 waves for the charged
and neutral modes.  These waves were included in the high-wave
set but not in the low-wave set.
}\label{leaks}
\end{figure}

One source of leakage into the exotic wave for the
PWA carried out with the low-wave set is the
$\pi_2(1670)$. 
In Figure~\ref{leaks} the coherent sum of the 
intensities for the 
$2^{-+}0^+ \ F$-wave $\rho\pi$,
$2^{-+}1^+ \ F$-wave $\rho\pi$ and $2^{-+}1^+ \ P$-wave $\rho\pi$ 
waves for the charged
and neutral modes are shown.  These waves are allowed decay modes of the $\pi_2(1670)$.  A clear peak is seen at the $\pi_2(1670)$ mass, the same mass
region where the purported exotic meson was seen 
 using the low-wave set.  The three waves shown in
Figure~\ref{leaks} were not included in the low-wave set.

\begin{figure}  
\centerline{\epsfig{file= 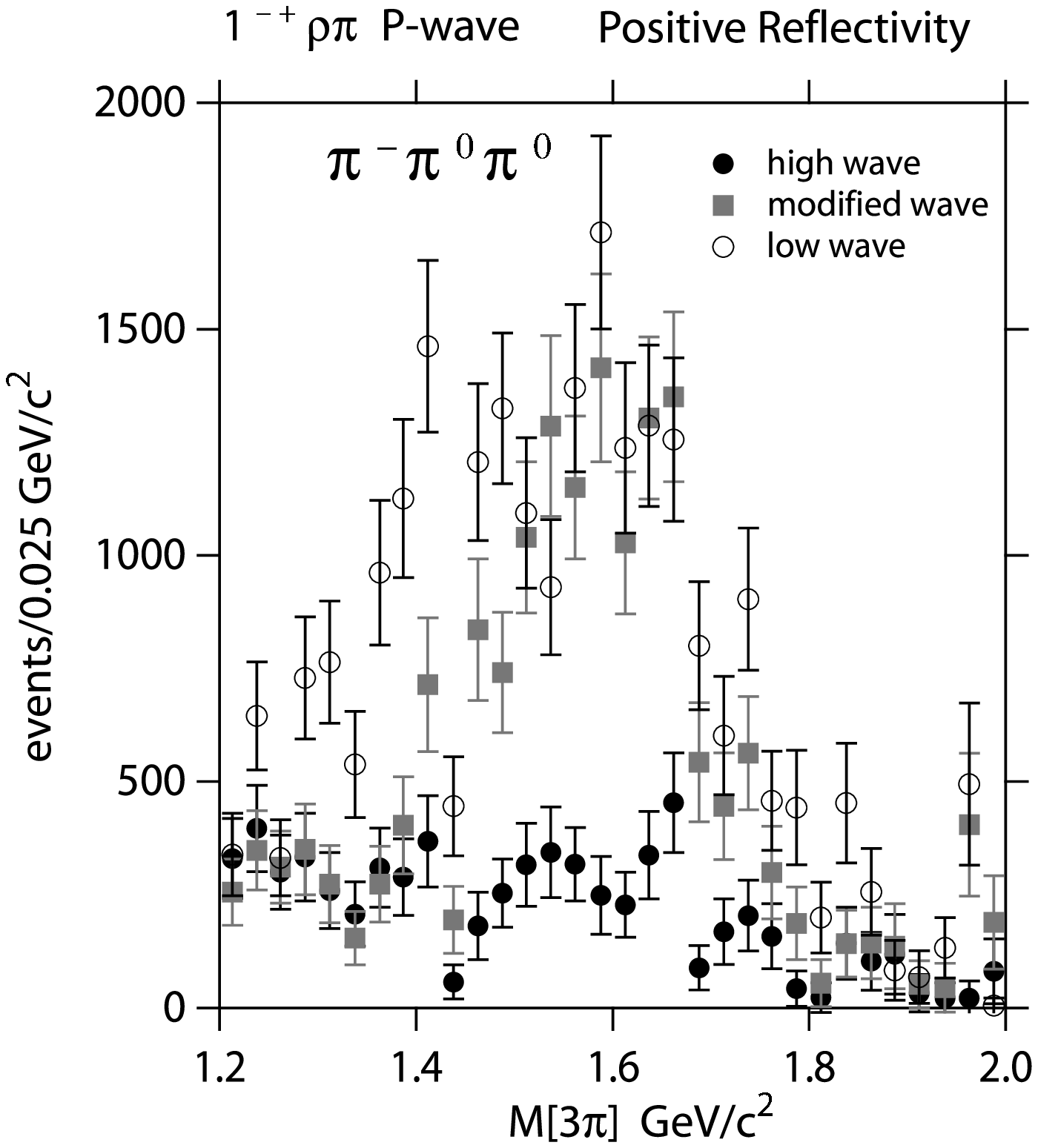,width=0.7\columnwidth}}
\caption{
The positive reflectivity $1^{-+} \ P$-wave $\rho\pi$ intensity for the neutral mode
for the high wave set (filled circles), the modified high wave set (filled
squares) and the low wave set (open circles).  In the modified high wave
set the two $2^{-+} \ F$-wave $\rho\pi$ waves 
and the $2^{-+}1^+ \ P$-wave $\rho\pi$ wave were removed.
}\label{neut_mod}
\end{figure}

\begin{figure}  
\centerline{\epsfig{file= 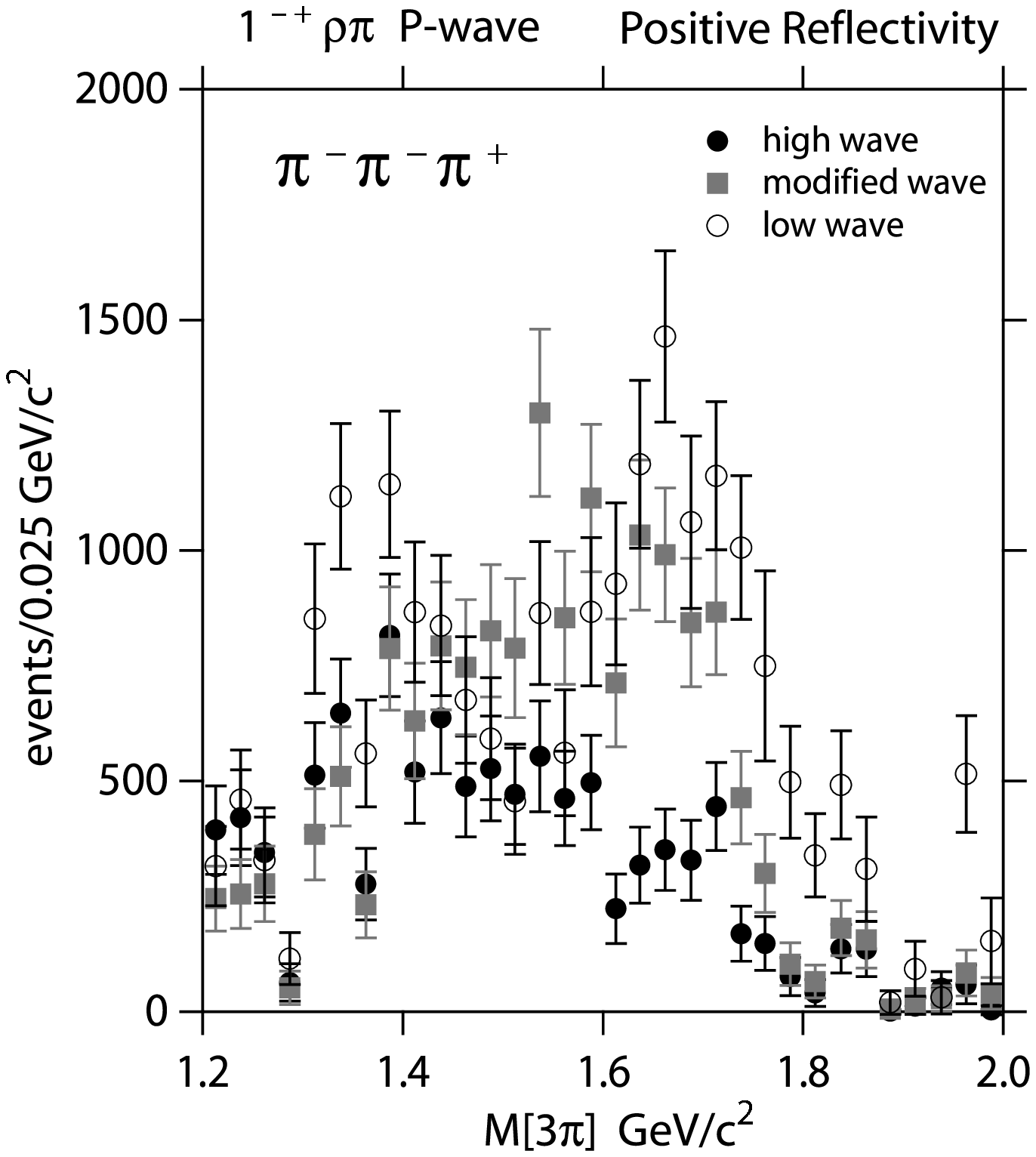,width=0.7\columnwidth}}
\caption{
The positive reflectivity $1^{-+} \ P$-wave $\rho\pi$ intensity for the charged mode
for the high wave set (filled circles), the modified high wave set (filled
squares) and the low wave set (open circles).  In the modified high wave
set the two $2^{-+} \ F$-wave $\rho\pi$ waves 
and the $2^{-+}1^+ \ P$-wave $\rho\pi$ wave were removed.
}\label{chrgd_mod}
\end{figure}

The PWA fit was repeated for the high wave set but with the 
$2^{-+}0^+ \ F$-wave $\rho\pi$,
 $2^{-+}1^+ \ F$-wave $\rho\pi$  and  $2^{-+}1^+ \ P$-wave $\rho\pi$
 waves removed.  
The resulting $1^{-+} \ P$-wave $\rho\pi$ intensities
 for the neutral mode for 
this modified high wave set fit, along with the unmodified high wave
set and low wave set, are shown in Figures~\ref{neut_mod}.
The corresponding results for the charged mode are shown
in Figure~\ref{chrgd_mod}.
The agreement between the modified high wave set and the low
wave set shows that the $2^{-+} \ F$-wave $\rho\pi$ decays and the
$2^{-+}1^+ \ P$-wave $\rho\pi$  decay of
the $\pi_2$ are the source of leakage leading to the observed
exotic peak in the low wave set.

\subsection{Evidence for leakage of the $a_4(2040)$}

 \begin{figure}  
\centerline{\epsfig{file= 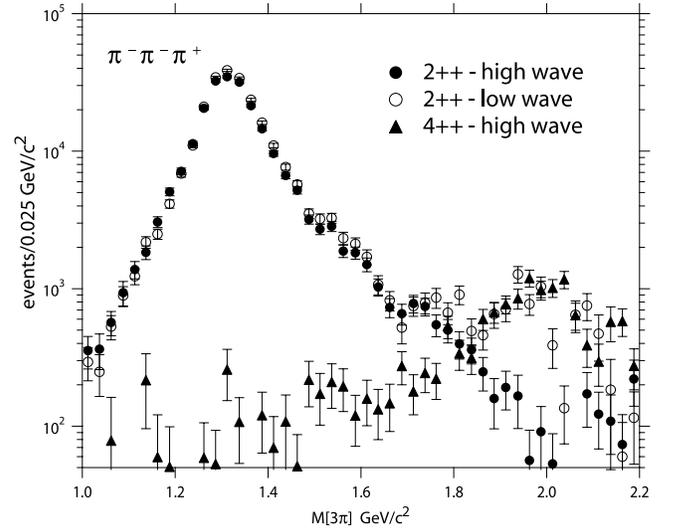,width=0.98\columnwidth}}
\caption{
The $2^{++}$ intensity for charged $3\pi$ mode for the high wave
set (filled circles) and low wave set (open circles).  The $4^{++}$
intensity (filled triangles) for the high wave set is also shown.
}\label{leakage_2_4}
\end{figure}

 A similar phenomenon is observed in the $2^{++}$ wave in the
low wave set where the $4^{++}0^+ \ G$-wave $\rho\pi$ is not included.  As
discussed above, the $4^{++}0^+ \ G$-wave $\rho\pi$ in the high wave set 
clearly shows the  $a_4(2040)$.  Figure~\ref{leakage_2_4}
shows a semi-log plot of the 
the $2^{++}$ intensity for charged $3\pi$ mode for the high wave
set (filled circles) and low wave set (open circles).  The $4^{++}$
intensity  for the high wave set is also shown (filled triangles).    The $a_2(1320)$ is dominant
in the  $2^{++}$ intensity in both high and low wave sets.  The
$2^{++}$ intensity for the low wave set shows a resonance-like structure
near 2~GeV/$c^2$ in low wave set but not the high wave set.  However
the  $2^{++}$ low wave set intensity agrees well with the $4^{++}$
high wave set intensity.  In this case, leaving out the $4^{++}$ in the 
low wave set forces a enhancement at $\sim2$~GeV/$c^2$
in the $2^{++}$ wave.

\subsection{Further discussion of the significance of the exotic wave}

\begin{figure}  
\centerline{\epsfig{file= 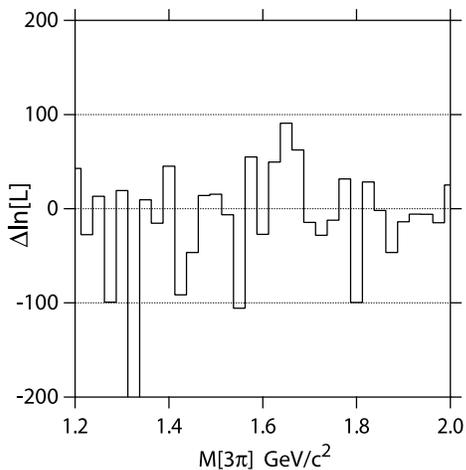,width=0.7\columnwidth}}
\caption{
The change in $ \ln {\cal {L}} $ resulting from the removal
of the $ 1^{-+}1^{+}(\rho \pi)$P-wave partial wave. See the text for
the sign convention and discussion.
}\label{exot_dlnl}
\end{figure}

The contribution to the $ \ln {\cal {L}}$ of the positive reflectivity exotic partial
wave was examined by removing this wave from the high wave set and comparing
the resulting likelihood with that of the original high wave PWA.
Figure \ref{exot_dlnl} shows the resulting change in likelihood.
The sign convention of this difference is such that 
positive (negative) values indicate inclusion of this partial wave improved 
(degraded) the fit quality.
As can be seen from figure \ref{exot_dlnl} there is no region of
$ 3 \pi $ effective mass where this partial wave is consistently
required by the fit.
This behavior is seen in all t-bins presented here.

\subsection{$t$ dependence studies}

\subsubsection{\label{sec:exotics}The exotic wave}

 \begin{figure}  
\centerline{\epsfig{file= 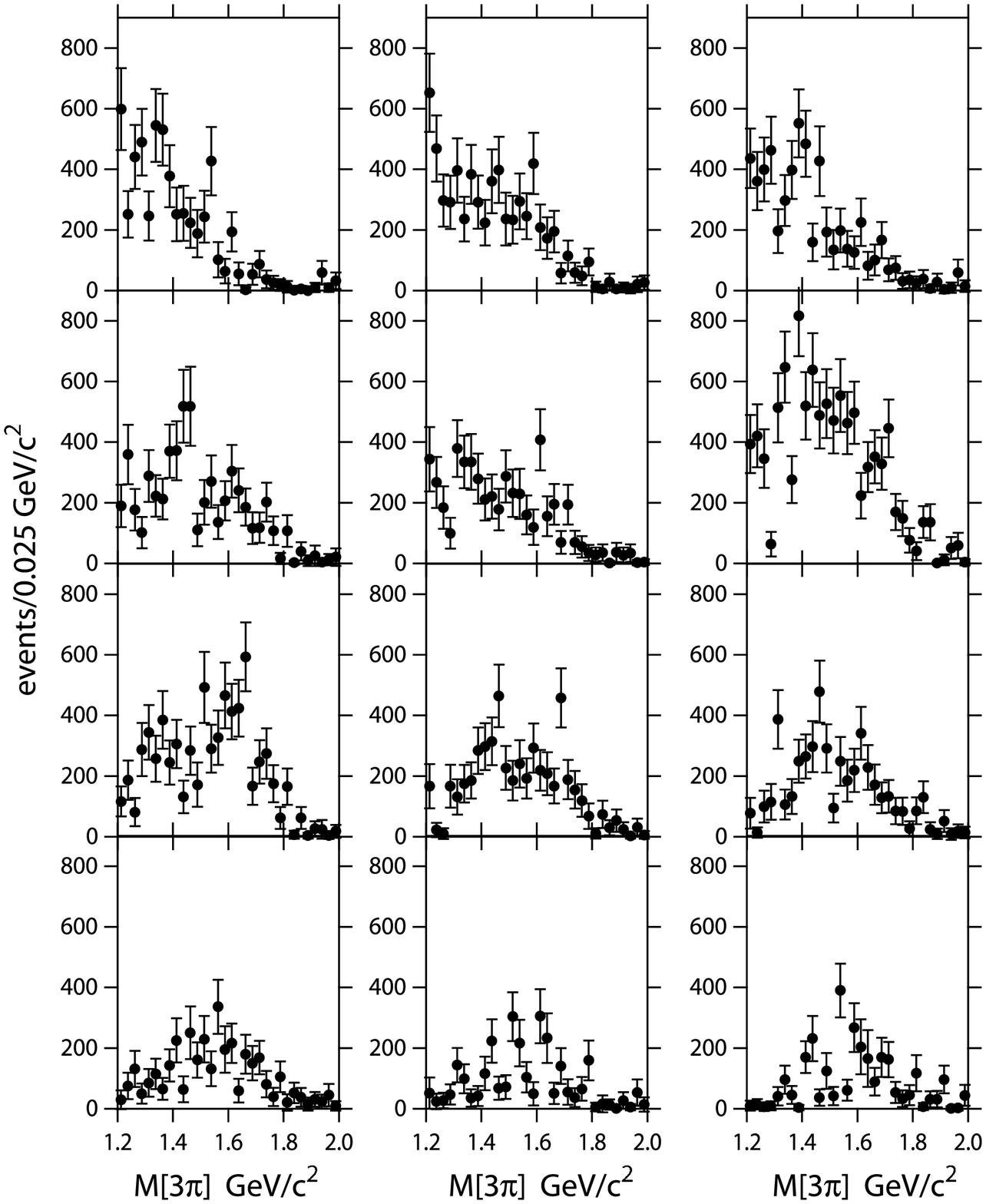,width=1.0\columnwidth}}
\caption{
Variation of the $1^{-+}1^+ \ P$-wave $\rho\pi$ intensity for the charged
mode for the 12 $t$-bins
of Table~\ref{tbins}
 increasing in $|t|$ from left to right and 
top to bottom. 
}\label{exoticw}
\end{figure}

 \begin{figure}  
\centerline{\epsfig{file= 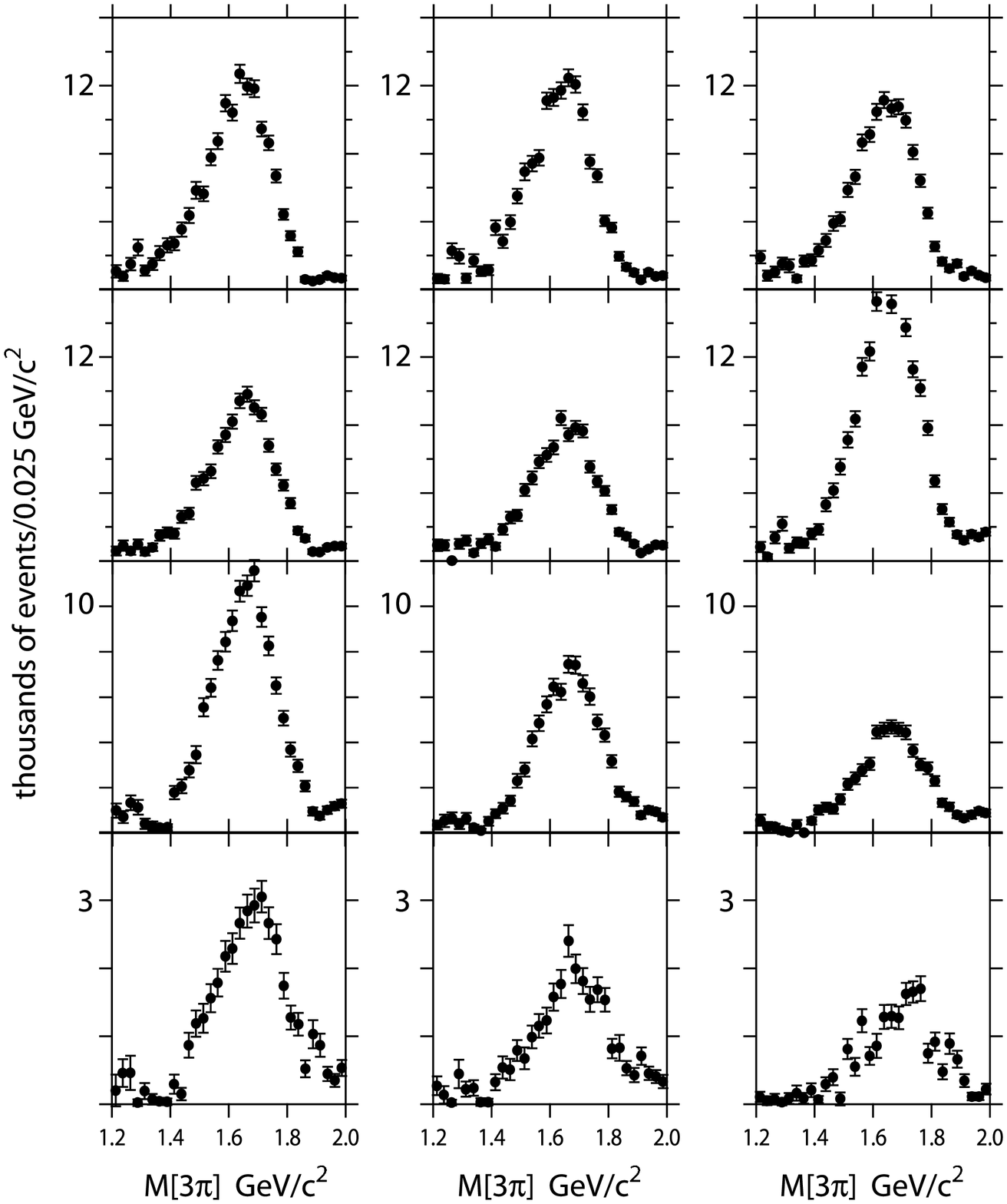,width=1.0\columnwidth}}
\caption{
Variation of the $2^{-+}0^+ \ S$-wave $f_2\pi$ intensity for the charged
mode for the 12 $t$-bins
of Table~\ref{tbins},
 increasing in $|t|$ from left to right and 
top to bottom.
}\label{pi2t}
\end{figure}

 \begin{figure}  
\centerline{\epsfig{file= 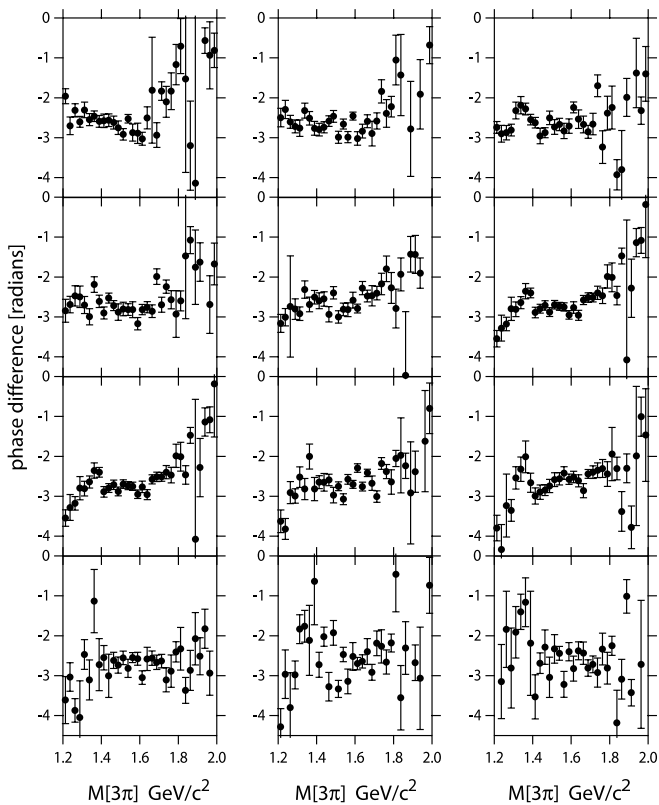,width=1.0\columnwidth}}
\caption{
Variation of the phase difference between the 
$1^{-+}$ and $2^{-+}$ amplitudes,  whose intensities are
shown in Figures~\ref{exoticw} and \ref{pi2t} respectively,
for the charged
mode for the 12 $t$-bins of Table~\ref{tbins}, increasing in $|t|$ from left to right and 
top to bottom.
}\label{phasex}
\end{figure}

 \begin{figure}  
\centerline{\epsfig{file= 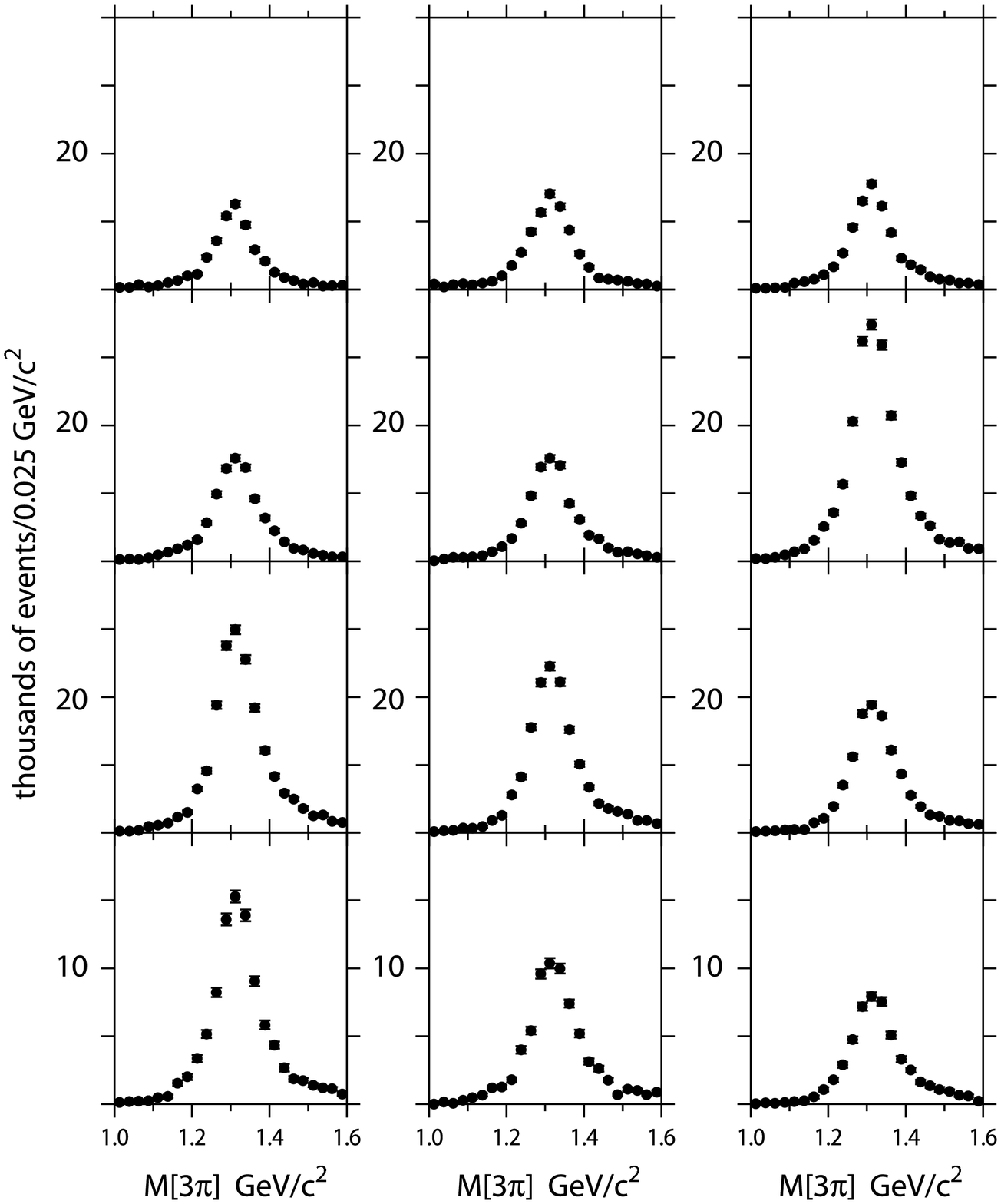,width=1.0\columnwidth}}
\caption{
Variation of the $2^{++}1^+ \ D$-wave $\rho\pi$ intensity for the charged
mode for the 12 $t$-bins of Table~\ref{tbins},
 increasing in $|t|$ from left to right and 
top to bottom.
}\label{plot_a2}
\end{figure}

 \begin{figure}  
\centerline{\epsfig{file= 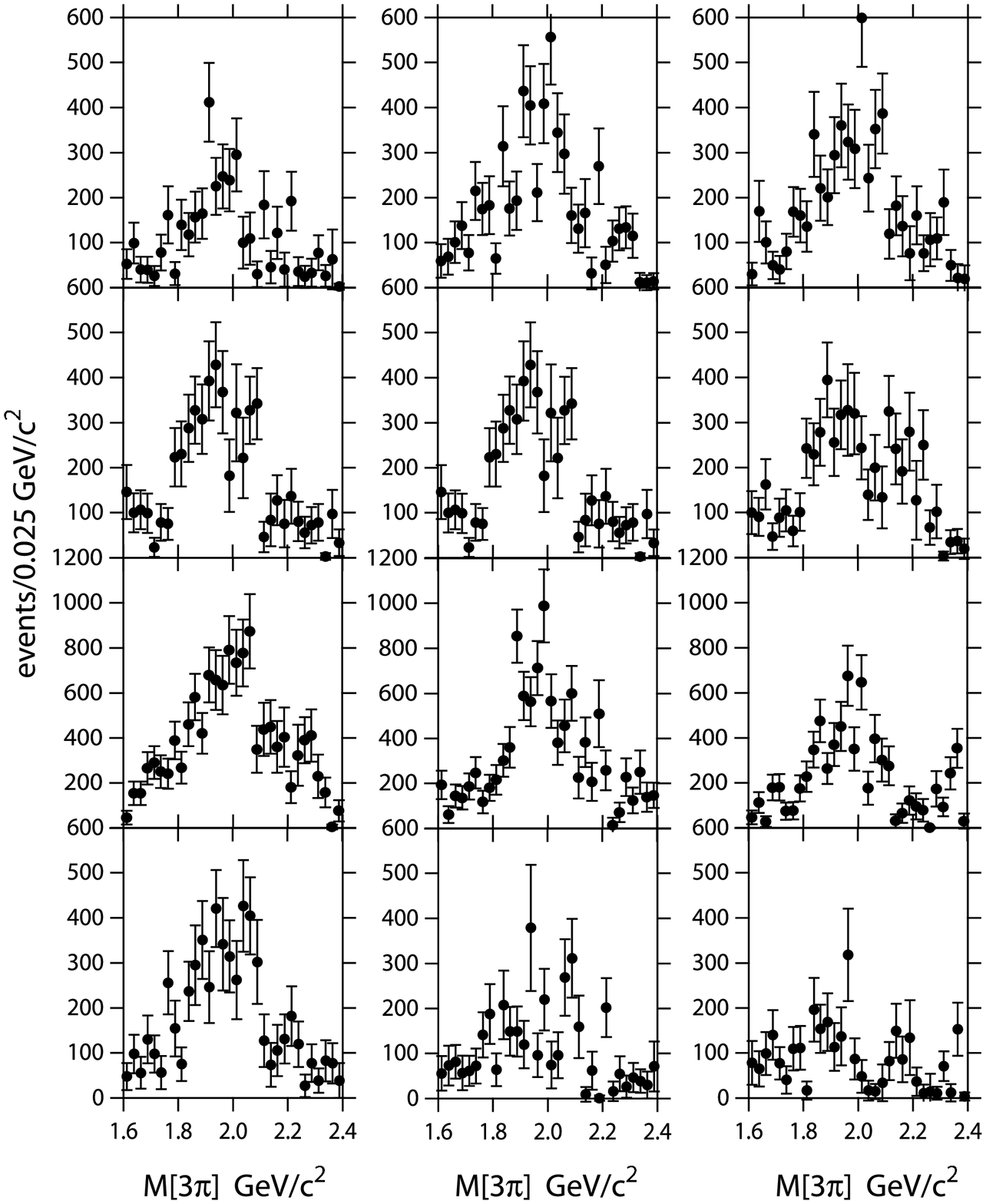,width=1.0\columnwidth}}
\caption{
Variation of the $4^{++}0^+ \ G$-wave $\rho\pi$ intensity for the charged
mode for the 12 $t$-bins of Table~\ref{tbins},
increasing in $|t|$ from left to right and 
top to bottom.
}\label{plot_a4}
\end{figure}

We examined the stability of the exotic wave intensity
and the intensity of other waves
across the 12 $t$-bins shown in Table~\ref{tbins}
 for the high wave fit in the charged $3\pi$. 
 As noted in Table~\ref{tbins}, the width of the $|t|$ bin for the first 
five bins is 0.02~(GeV/$c$)$^2$ and 0.05~(GeV/$c$)$^2$ for the
other bins.
Figures~\ref{exoticw}, \ref{pi2t} and \ref{phasex} show the 
variation of the $1^{-+}1^+ \ P$-wave $\rho\pi$ intensity,
the $2^{-+}0^+ \ S$-wave $f_2\pi$ intensity and the
phase difference between the 
$1^{-+}$ and $2^{-+}$ amplitudes 
 for the charged $3\pi$
mode for the 12 $t$-bins. 
The $2^{-+}0^+ \ S$-wave $f_2\pi$ intensity
is stable across the 12 $t$ bins. 
We note that although the $1^{-+}1^+ \ P$-wave $\rho\pi$ intensity
may indicate an enhancement in the vicinity of 
$\sim1.6$~GeV/$c^2$ for some of the higher $t$ bins
in Figure~\ref{exoticw}, the corresponding phase difference
shown in Figure~\ref{phasex} is not consistent with resonant
behavior. Indeed  the $2^{-+}0^+ \ S$-wave $f_2\pi$ and
$1^{-+}1^+ \ P$-wave $\rho\pi$ are phase-locked in the 
1.6~GeV/$c^2$ region.  One possible explanation is production of 
an exotic resonance with the same mass, width and phase motion as the
$\pi_2(1670)$.  Another is some residual leakage of the $\pi_2(1670)$
in the exotic wave  in the high-wave set.

\subsubsection{The $2^{++}$ and $4^{++}$ waves}

\begin{figure}  
\centerline{\epsfig{file= 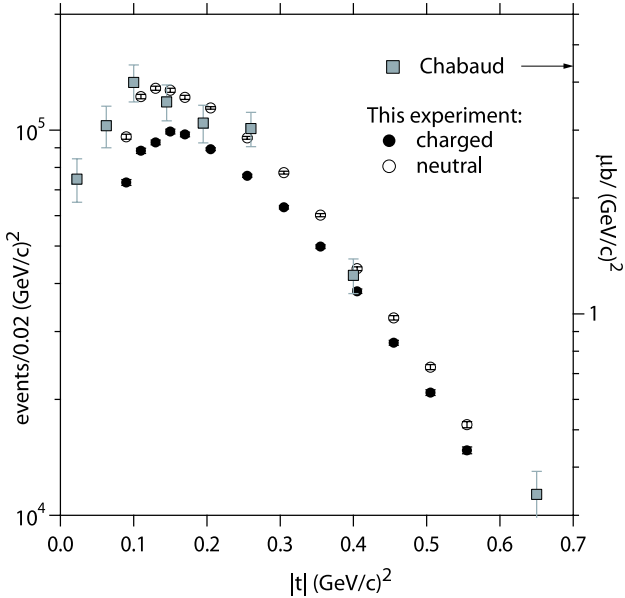,width=0.95\columnwidth}}
\caption{
$t$ dependence of the $a_2(1320)$ as measured in this experiment
for the charged mode (filled circles) and neutral mode (open
circles) along with measurements of V.~Chabaud {\it et. al.} \cite{Chabaud} 
of $\pi^-p \to a_2^-(1320)p$ at 18.8~GeV/$c$.
}\label{dsdt_a2}
\end{figure}

The 
$2^{++}1^+ \ D$-wave $\rho\pi$ intensity (shown in 
Figure~\ref{plot_a2}) and the 
 $4^{++}0^+ \ G$-wave $\rho\pi$ intensity (shown in 
Figure~\ref{plot_a4}) are shown for the 12 $t$ bins.
For completeness, in Figure~\ref{dsdt_a2},
we compare the $t$ dependence of the 
$2^{++}1^+ \ D$-wave $\rho\pi$ intensity, dominated by the $a_2(1320)$,
in both $3\pi$ modes in this experiment, with other measurements
of $\pi^-p \to a_2^-(1320)p$ at 18.8~GeV/$c$\cite{Chabaud}.

\section{\label{sec:systematics}Studies of systematics}

\subsection{Effects of varying data selection cuts}

Studies were carried out to determine the effect on the PWA of varying the
parameters that define the selection criteria as described in section~\ref{sec:data}.
Different criteria were found to effect the results in different ways but no
significant change in our conclusions was observed. The effects of
these perturbations on the analysis are described individually below.

The beam hole cut is nominally $ 2.5 \sigma $.
Values of $ 1.5 \sigma $ and $ 3.5 \sigma $
systematically increased and decreased the intensity of the $ 1^{++} \rho
\pi$ $S$-wave
below 1.2~GeV/$c^2$ but left the mass region above 1.2~GeV$/c^2$
indistinguishable from the unperturbed analysis. The changes
observed
were approximately  5\%
of the
intensity. Conclusions regarding the low mass behavior of this partial wave
should take into account this systematic effect. Other partial waves were
unaffected.

The cut to remove the $ \Delta^{++}$ in the charged sample affected the
data only above 2.0 GeV/$c^2$.  Removing this cut from the analysis
increased the intensity of the $ 2^{-+}0^+(f_2 \pi) $ 
$S$-wave by approximately twice
its error between 2.0 and 2.4  GeV/$c^2 $. The systematic effect on other
partial waves is insignificant when compared with their statistical errors.

The effect of removing the DEA cut was similar to using a less restrictive
value for the beam hole cut. Small, but statistically significant,
increases in the intensity of the $ 2^{-+}0^+(f_2 \pi) $ $S$-wave
were observed
between 1.2 and 1.6  GeV/$c^2 $. 

Increasing (decreasing) the confidence level cut to 0.3 (0.1) changed the
observed intensities everywhere by a mass independent scale factor.
The same effect was observed for the CSI cut.  Removing this cut increased
all intensities by a factor of 1.66 but left their shapes unchanged.

In summary,  we have investigated the effect of perturbing the selection
criteria summarized in Tables~\ref{cutsn} and \ref{cutsc} on the results
of the partial wave fit and have shown that our results do not depend on the
 specific values used in the criteria.

\subsection{ $S$-wave parametrization}
We used the isobar model, following the same methodology employed
in  references~\cite{Adams98,Chung02}.
This model uses  two-body amplitudes for the
 $\pi\pi$ system.   The $S=1$ isovector  and $S=2$  isoscalar amplitudes were parametrized using Breit-Wigner amplitudes. There are various $S=0$, isoscalar  parametrizations available in the literature.  In general the  $\pi\pi$ amplitude is dominated by a broad $\sigma$ meson, the $f_0(980)$ and several resonances above $1\mbox{ GeV}/c^2$, depending on the specific  parametrization. The results presented in this paper were all based on the parametrization used in
 references~\cite{Adams98,Chung02}  . This is based on the model of ~\cite{AMP} with the possibility of direct $f_0(980)$ production;  the $f_0(980)$ was subtracted from the elastic $\pi\pi$ amplitude and added as a Breit-Wigner resonance.  We have also studied other parametrizations, {\it e.g.} based on (1)~$N/D$ dispersion relations, (2)~explicit inclusions of the  $K{\bar K}$ channel and/or 
(3)~including direct production of scalar resonances above $1 \mbox{ GeV}/c^2$~\cite{oset,oset-nd}. 
All give qualitatively identical results.

\section{\label{sec:conclude}Conclusions}

A partial wave analysis of the $\pi^- \pi^-\pi^+$ and
$\pi^- \pi^0\pi^0$ systems has been presented using a data set
whose statistics exceeds published analyses of the $3\pi$ systems.
The
 $a_1(1260)$, $a_2(1320)$, $\pi_2(1670)$
and $a_4(2040)$  mesons 
were observed.
The data provide no
evidence  for an exotic $J^{PC}=1^{-+}$ meson
 as had been reported earlier using
a smaller sample of the $\pi^- \pi^-\pi^+$ mode alone \cite{Adams98,Chung02}. 
This analysis included the study of criteria for defining a set of partial waves sufficient
to describe the data including a study of the effect on the fit quality by
removing individual waves and also a comparison of observed moments of
the data with moments computed from PWA solutions.  The wave set
thus established contained more waves than were used in the analysis
reporting evidence for the exotic meson.  Our studies indicate that leaving
out partial waves corresponding to decay modes of the $\pi_2(1670)$
lead to a false enhancement in the exotic $1^{-+}$ wave.  
This analysis and the analysis of references~\cite{Adams98,Chung02} were
based on the isobar model.  The isobar model is known
to suffer some limitations, as has been pointed by others \cite{Ascoli73,Ascoli74}.
The so-called {\it Deck mechanism} \cite{Ascoli73,Ascoli74} is particularly
relevant to the production of the $3\pi$ system and is being applied to these
data.  Results will be presented in a forthcoming paper.

\begin{acknowledgments}
The authors thank Jozef Dudek, Geoffrey Fox and Matt Shepherd
for enlightening discussions and
for their careful read and critique of this paper.
This work was supported by grants from the United States Department of
Energy
(DE-FG-91ER40661/Task~D and DE-FG0287ER40365),
 the National Science
Foundation (EIA-0116050) for AVID, the Russian Ministry of
Science and Technology and Russian Ministry of Science and Education.
\end{acknowledgments}

\appendix

\section{\label{sec:online} Technical notes and  online PWA results}

The authors of this paper have made available technical notes and PWA
results online \cite{3piE852}.  The notes include (1)~details of the data selection;
(2)~a comparison of the software used in this analysis with that used
in the analysis of reference~\cite{Chung02};  (3)~the Monte Carlo details 
for this analysis;  (4)~a full description of the PWA formalism;  (5)~the moments
method technique and definition of angles;  (6)~a study of background contributions
to partial wave intensities; (7)~ stability of PWA results against variations in 
data selection criteria; (8)~ acceptance and resolution; (9)~the determination
of the minimal partial wave set and (10)~$S$-wave parametrization.
 In addition, PWA results (amplitudes and phases)
are available for 12~bins in $t$ and for various wave sets.



\end{document}